\documentstyle[a4,12pt]{article}
\title{Symmetries of KdV and Loop Groups}
\author{Jeremy Schiff\\
     Department of Mathematics and Computer Science\\
     Bar Ilan University\\
     Ramat Gan 52900, Israel\\
     e-mail: schiff@math.biu.ac.il}
\date{May 1996}

\begin{document}
\maketitle
\begin{abstract}
A simple version of the Segal-Wilson map from the $SL(2,{\bf C})$ loop group
to a class of solutions of the KdV hierarchy is given, clarifying certain
aspects of this map. It is explained how the known symmetries, including
B\"acklund transformations, of KdV arise from simple, field independent, 
actions on the loop group. A variety of issues in understanding the 
algebraic structure of B\"acklund transformations are thus resolved.
\end{abstract}

\section{Prelude: Symmetries of KdV} \label{symms}

This section contains a list, which I believe is exhaustive, of known notions
of ``symmetry'' of the KdV hierarchy. One of the aims of this paper is to
obtain a unified, simple understanding of them.

\subsection{Translation Invariance}

The KdV hierarchy is a system of autonomous differential
equations on an infinite dimensional affine space, and hence an
infinite dimensional abelian group, associated with translation 
invariance, acts on the space of solutions.

\subsection{Scaling and Galilean Invariance}

Two further one parameter groups are known to act on the space of solutions,
the ${\bf C}^*$ action associated with scaling invariance (see, for example, 
\cite{DJ}, for the case of the  KdV equation), and the ${\bf C}$ action 
associated with Galilean invariance (see, again, \cite{DJ}, for the case of 
the KdV equation). The full group generated by transformations of these 
types and translations is easy to identify.

\subsection{Wahlquist-Estabrook B\"acklund Transformations}

Another well known, but less well understood, symmetry is the one parameter 
family of Backlund transformation (BTs) of Wahlquist and Estabrook \cite{WE}
(see \cite{DJ} sections 5.4.2 and 5.4.3 for a typical exposition). To 
implement a BT is in two ways harder than the simple transformations listed
above: first, implementing a BT involves solving a differential equation, 
and second, the form of this differential equation depends on the solution 
being transformed, i.e. the transformation is {\em field dependent}.
Note that together, these two problems mean it is not clear when
BTs can be implemented: it is necessary to check that the solution being
transformed has the properties
that guarantee global solvability of the required differential equation. 

When BTs can be applied, several remarkable algebraic properties em\-erge.
Because of an integration constant that appears in each application of
a BT, BTs do not map single solutions to single solutions, but rather the
image of a single solution is a {\em set} of solutions. Thus a BT 
is an operator mapping the space of sets of solutions of KdV
to itself; if the parameter in the BT is $\theta\in{\bf C}$, let us denote 
this operator as ${\cal O}_\theta$. The known algebraic properties
(assuming applicability of the BTs) are (1) commutativity \cite{WE}, (2) 
that the image of any set of solutions under the square of a BT (of
a given parameter value) includes the original set (this is evident 
from the equations defining the BT), and (3) the product of two 
BTs is {\em not} itself another BT. That is, 
\begin{eqnarray}
&\forall~\theta_1,\theta_2~~~~~{\cal O}_{\theta_1}{\cal O}_{\theta_2}=
      {\cal O}_{\theta_2}{\cal O}_{\theta_1}& \nonumber \\
&\forall~\theta~~~~~I\subset{\cal O}_{\theta}^2  & \label{alg-relns}\\
&\not\exists~\theta_1,\theta_2,\theta_3~~{\rm s.t.}~~
{\cal O}_{\theta_1}{\cal O}_{\theta_2}\neq{\cal O}_{\theta_3}.&
     \nonumber 
\end{eqnarray}
(In the second of these relations, I use the notation that for operators
$A,B$ on sets, $A\subset B$ implies the image of any set under $A$ is a 
subset of the image of that set under $B$, and $I$ denotes the identity 
operator on sets.)

In addition to the problem of determining to which sets of solutions BTs can 
be applied, there is a need to understand the origins of this strange notion 
of symmetry, and how it interacts with the other symmetries listed in this 
section. The results of this paper go some way towards resolving these last 
two questions; in particular I will give in Sec.~\ref{WEBT} what I consider 
to be the correct algebraic framework within which to consider BTs, and the 
results of Sec.~\ref{expl-syms} give a field independent realization of all 
the symmetries of KdV listed in this section, making their interrelationships 
straightforward. With regard to the question of applicability of BTs, the 
paper \cite{GeS} studies when the BT ${\cal O}_0$ can be implemented on a 
solution satisfying ``reasonable'' analyticity conditions while retaining 
such conditions. A sufficient condition for this is found: that the spectrum 
of the Schr\"odinger operator associated with the starting solution have no
negative eigenvalues. The solutions of KdV studied here
(which include certain singular solutions) all admit the BT ${\cal O}_0$, 
and also BTs of other sufficiently low parameter values, while 
staying in the class of solutions under study. It seems, however, that the
class only includes single soliton solutions that are sufficiently broad,
shallow and slow (recall that all these properties are determined by a single
parameter), and thus the class of solutions is not sufficiently large to
make the question of when a BT can or cannot be applied, while staying 
within the class, of real interest.

\subsection{Galas B\"acklund Transformations}
\label{Galas}

In a relatively unknown recent work \cite{Galas}, Galas found a novel 
B\"acklund transformation for KdV. Since this is not well known, and 
presented only very briefly in \cite{Galas}, I give some details.
Combining Eqs.~(4) to (7) in \cite{Galas}, it emerges that the KdV equation
\begin{equation}
u_t+u_{xxx}+6uu_x=0
\end{equation}
is invariant under $u\mapsto u+2(\ln\tau)_{xx}$, where $\tau$
(which Galas calls $4+\epsilon p_3$) is related to $u$ by
\begin{equation}
u = \theta - \frac{1}{2}\left( \frac{\tau_{xxx}}{\tau_x} -
                      \frac{\tau_{xx}^2}{2\tau_x^2}  \right) 
  = \theta - \frac{(\sqrt{\tau_x})_{xx}}{\sqrt{\tau_x}}
\label{GBT1} \end{equation}
and 
\begin{equation}
\frac{\tau_t}{\tau_x} = -6\theta -\left( \frac{\tau_{xxx}}{\tau_x} -
                      \frac{3\tau_{xx}^2}{2\tau_x^2}  \right). 
\label{GBT2} \end{equation}
For contrast, the standard BT is $u\mapsto u+2(\ln\tau)_{xx}$, 
where $\tau$ is related to $u$ by
\begin{equation}
u = \theta - \frac{\tau_{xx}}{\tau}
\label{f1}\end{equation}
and 
\begin{equation}
\frac{\tau_t}{\tau_x} = -6\theta -\left( \frac{\tau_{xxx}}{\tau_x} -
                      \frac{3\tau_{xx}}{\tau}  \right). 
\label{f2}\end{equation}
Galas finds the infinitesimal version of his BT as well, and
shows that it generates one soliton solutions from the trivial solution 
$u=0$. In fact it can generate other interesting solutions from $u=0$; it 
is straightforward to check that
\begin{equation}
\tau=(x-12k^2t) + \frac{\sinh(2k(x-4k^2t))}{2k}
\end{equation}
solves Eqs. (\ref{GBT1}) and (\ref{GBT2}) with $u=0$ and $\theta=k^2$. This
gives an interesting singular solution of KdV, which should presumably be 
regarded as a nonlinear superposition of the rational solution 
$u=2(\ln\tau)_{xx}$, $\tau=x$, with  the ``singular soliton'' solution 
$u=2(\ln\tau)_{xx}$, $\tau=\sinh(2k(x-16k^2t))$ (at least up to Galilean 
transformations). Note, however, that like both these solutions, for any real 
value of $t$ the new solution is only singular for one real value of $x$
(and has a double pole there); is this sense the notion of ``superposition''
is inappropriate. Such mixed rational-solitonic solutions have been 
studied in the literature \cite{ratsol}.

Algebraic properties of this BT have not as of yet been given; in 
Sec.~\ref{GBT} I will prove commutativity, at least within a limited
class of solutions. The possible limitations on applicability, the fact 
that one BT creates a family of solutions out of a single one, and the fact 
that this transformation is field dependent, all apply to this BT as they do 
to the standard one.

\subsection{The Hierarchy of Infinitesimal Symmetries Generalizing 
Scaling and Galilean Invariance}
\label{infsys}

It is of course possible to write generators of translation, scaling and 
Galilean transformations. The generators of the translation transformations 
form a {\em hierarchy}; they are related to each other by application of an 
operator, with very special properties, known as the recursion operator
(see, for example \cite{Olver}). 
The ``lowest order'' translation is generated by the recursion operator out 
of the generator of the trivial symmetry, i.e. the symmetry that leaves a
solution invariant. The generators commute, reflecting the abelian nature of
the translation group. 

It turns out that the generator of scaling transformations can be obtained
from the generator of Galilean transformations by application of the 
recursion operator. And, furthermore, a hierarchy of symmetry generators
can be produced by repeated application of the recursion operator to the 
generator of Galilean transformations. Let
us denote the generators of the translation symmetries by $l_n$, 
$n=0,1,2,\ldots$, (where it is understood that $l_{n+1}$ is obtained by
the action of the recursion operator on $l_n$, and $l_0$ is the lowest order
translation generator), and the generators in
the new hierarchy by $m_n$, $n=-1,0,1,\dots$ (where again it is understood
that $m_{n+1}$ is obtained by the action of the recursion operator on
$m_n$, and $m_{-1}$ is the generator of Galilean transformations). The
algebra obeyed by these infinitesimal transformations is then found to be
\begin{eqnarray}
[l_r,l_s] &=& 0 \nonumber \\
{}[m_r,m_s] &=& (s-r)m_{r+s} \\
{}[m_r,l_s] &=& (s+{\textstyle{\frac12}})l_{r+s} \nonumber 
\label{infsymsalg} \end{eqnarray}
(where in the last relation it is understood that $l_{-1}=0$). It is not
currently known how to exponentiate the generators $m_n$, $n>0$; of
particular interest is the generator $m_1$, since $m_{-1},m_0,m_1$ form
a closed $sl(2)$ subalgebra.

The history of the $m_n$ symmetries is a little unclear to me. The 
earliest reference I am aware of is \cite{IS}. They are 
exploited in \cite{hidsym}, and a reference is given to a
preprint by the authors of \cite{CLL}; in \cite{CLL} these infinitesimal 
symmetries were discovered for the KP hierarchy (although all solutions of 
KdV are solutions of KP, it does not follow that symmetries of KP can be 
restricted to symmetries of KdV; in this case they can, as has also been
illustrated in \cite{ASvM}). The symmetries are discussed 
further in \cite{Kor}. In addition,
there has been some discussion \cite{GH1} of how the $m_n$ hierarchy
can be extended to define symmetry generators $m_{-2},m_{-3},\dots$ by
application of the inverse of the recursion operator. These 
results will be reproduced later in the paper 
(and also results of \cite{GH2}
showing the existence of further infinitesimal symmetries that leave 
the KdV solution invariant but act on certain ``prepotentials'' 
associated with the solution).

\subsection{Zakharov-Shabat Dressing Transformations}
\label{dress}

Somewhat remote from all the above results, but, in a sense that will
emerge, inclusive of all of them, are the ``dressing transformations''
of Zakharov and Shabat (see \cite{Wilson} for a concise explanation
in the case of the modified KdV equation). The KdV hierarchy can be
interpreted as the consistency condition for a certain homogeneous
linear differential system. A solution of this linear system gives rise 
to a solution of the hierarchy, but there are many solutions of the 
linear system corresponding to any solution of the hierarchy. 
Dressing transformations should be thought of as an action of the
$SL(2,{\bf C})$ loop group on solutions of the linear system. Unfortunately,
taking two solutions of the linear system corresponding to the 
{\em same} solution of the hierarchy, and acting on them with the {\em same}
dressing transformation, gives, in general, two solutions of the 
linear system corresponding to {\em different} solutions of the hierarchy.
Thus there is no guarantee that  dressing transformations
give rise to a sensible notion of 
symmetry on the hierarchy itself. It turns out that there is a dense
subset in the loop group for which dressing transformations can be 
interpreted as B\"acklund transformations (i.e. as maps that take a 
single solution to a family of solutions of finite dimensionality); 
the Galas BTs arise precisely this way. The standard BT, however, can only 
be expressed as a field {\em dependent} dressing transformation. 

\section{Aims and Methods}
\label{am}

This paper has two main aims. One, as stated above, is to obtain a unified
and simplified understanding of symmetries of KdV. I will do this by
exploiting a cornerstone of KdV theory, the {\em Segal-Wilson correspondence}.
The other aim of this paper is to give a simplified reformulation of
this correspondence, clarifying several issues, both conceptual and 
technical.

In \cite{SW}, Segal and Wilson gave a construction associating a solution of 
the KdV hierarchy with each point in a certain infinite dimensional 
grassmanian. The grassmanian is a homogeneous space of the $SL(2,{\bf C})$ 
loop group, which I denote $G$, a quotient of $G$ by the 
subgroup of loops which are boundary values of analytic loops on the unit 
disc; I denote this subgroup $G^+$. Throughout this paper I regard 
$G$ as the set of $2\times 2$ matrices, with unit determinant, and entries 
Laurent series in a parameter $\lambda$, convergent for 
$\vert\lambda\vert=1$, and defining a smooth map from the circle
$\vert\lambda\vert=1$ to $SL(2,{\bf C})$; thus $G^+$ is the subgroup for 
which entries are power series, which, since they are
convergent for $\vert\lambda\vert=1$, define analytic functions in
$\vert\lambda\vert< 1$. The main paper of Segal and Wilson \cite{SW} does 
not really exploit the description of the grassmanian as the quotient 
$G/G^+$; however in other papers, Wilson \cite{Wilson} (actually for the case
of the modified KdV hierarchy) gives an explicit map from elements of $G$ to
solutions of the hierarchy, exploiting the Birkhoff factorization of $G$.
(Writing $G^-$ for the subgroup of loops that are boundary values of loops 
analytic in $\vert\lambda\vert>1$ and which reduce to the identity at 
$\lambda=\infty$, the Birkhoff factorization theorem states that for elements 
$g$ in a dense, open subset of $G$, there exists a unique factorization 
$g=g_-^{-1}\cdot g_+$, where  $g_-\in G^-$, $g_+\in G^+$. The product map 
$G^-\times G^+\rightarrow G$ defined by $(g_-,g_+)\mapsto g_-^{-1}g_+$ is a 
diffeomorphism from $G^-\times G^+$ to its image. See \cite{loopgroups}.)
The solution of the hierarchy obtained by Wilson's map from an element
$g\in G$ is unchanged by right multiplication $g\mapsto g\cdot h$, 
$h\in G^+$; thus the map actually defines a map from $G/G^+$ to solutions
of the hierarchy.

The reformulation of this map that I will give is inspired by Mulase's 
results for the KP hierarchy \cite{Mulase} (which I had the good fortune to
hear Mulase lecture on in 1989). Mulase emphasizes that integrable systems
are really  linear systems in disguise. Of course, for KdV this is 
well-known: the invertible scattering transform converts the nonlinear
flow for the function satisfying the KdV hierarchy to a linear flow
for associated scattering data. I will consider a simple linear flow
on the loop group $G$, whose solution is determined uniquely by an initial
value, i.e. some element of $G$. A flow on $G$ induces, via Birkhoff
factorization, and assuming the flow does not leave the relevant dense
open subset of $G$, flows on $G^+$ and $G^-$, and there is no reason
why these flows should be linear. The corresponding flow on
$G^-$ turns out to be, more or less, the KdV hierarchy! The orbit on $G^-$   
is unaffected by right multiplication of the initial value of the flow on
$G$ by an element of $G^+$ and thus Wilson's map from $G/G^+$ to
solutions of KdV is recovered. 

The reader may be itching to know what I meant by the phrase ``more or
less'' in the above paragraph. In fact the $G^-$ flow contains somewhat
more than the KdV hierarchy. In particular I will identify a further
infinite dimensional abelian subgroup $H$ of $G$, that, acting in an 
appropriate way on the initial value of the $G$ flow, leaves the associated
KdV solution invariant, even though the $G^-$ flow is not invariant. 
It thus turns out that the Wilson map is a map from the double coset space
$H\backslash G/G^+$ to solutions of KdV; a study of the geometry of this
space would be very interesting. 

Returning for a moment from technical to conceptual issues, the approach
I present ``explains'' the role of the loop group in KdV theory: the loop 
group is the space of initial value data for the simple linear flow 
that KdV conceals.

Another technical issue that emerges is that there are two ways
to pass from the $G$ flow to the associated KdV solution, using either the 
$G^+$ or the $G^-$ flows as intermediaries. These both have an advantage and
a disadvantage: the $G^+$ flow is not invariant under the $G^+$ action that
leaves the KdV solution invariant, but is under the $H$ action, and 
vice-versa for the $G^-$ flow. For different computational purposes,
different approaches are preferable. Note the $G^+$ flow is also 
determined by a set of linear equations; this is the linear system referred
to in the discussion of dressing transformations in section \ref{dress}. 
In Wilson's works \cite{Wilson} only the $G^+$ flow is used; on the other
hand, in the work of Drinfeld and Sokolov \cite{DS}, the $G^-$ flow is used. 
The results presented here show the rather simple way
these approaches are related.

Armed with a firm grasp of the Segal-Wilson correspondence, 
the symmetries of KdV can be understood. The natural origin of symmetries
of KdV should be what will be loosely called ``symmetry actions'' on the loop
group. These include right and left multiplication by elements of $G$, other
automorphisms of $G$, and the induced action on (subgroups of) $G$ of
reparametrizations of $\lambda$. Unfortunately most of these actions do not 
descend to the double coset space. It turns out that there are some actions
that do (these give rise to genuine symmetries of KdV), and others that have
the property that they map individual cosets to a finite dimensional set
of cosets. These latter actions are precisely B\"acklund transformations!
In this way field independent symmetry actions on $G$ give 
rise to all the symmetries of KdV I have listed in section \ref{symms}. 
Field independence is an important point here; all of the symmetry
actions on $G$ that will be considered can be rewritten as, say, right
multiplications, but not necessarily field independent ones.

The contents of the remainder of this paper are as follows. In section
\ref{zc}, I discuss the zero curvature formulation of the KdV hierarchy,
and particularly a nonstandard gauge choice that will be important.
In section \ref{the-constr} I present the reformulation
of the Segal-Wilson correspondence, and, in section \ref{expl-syms} I
use it to show how field independent symmetry actions on $G$ induce all 
the symmetries of KdV presented in section \ref{symms}. Finally, in 
section \ref{concl}, I present a few open problems.

Two more comments are in order before concluding this introductory
section. First, the reader will have noticed that I am studying the 
KdV hierarchy without exploiting its embedding in the KP hierarchy. The
reason for this is quite simple --- although many of the known
properties of KdV are inherited from those of KP, it is not the case
that this must be so for all interesting properties of KdV, and in
particular, I am not certain whether the Galas BT has an analog for KP.
Second, the reader will find another presentation of a more
``group-theoretic'' approach to the Segal-Wilson correspondence in 
\cite{HSS}, which has some overlap with the ideas being presented here,
and shows that the ideas being presented here can be set in a very
general framework. The work of Mulase \cite{Mulase} is actually an 
example of the constructions of \cite{HSS}.

\section{A Nonstandard Gauge for the Zero Curvature Formulation of KdV}
\label{zc}

The KdV hierarchy is defined as follows. The sequence 
$P_n$, $n=0,1,2,\ldots$  of differential polynomials in $u(x)$ 
is defined by the recursion relation
\begin{equation}
\partial_x P_{n+1} = ({\textstyle{\frac14}}\partial_x^3 
 -  u\partial_x - \partial_x u) P_n  ~~~~~~n=0,1,2,\ldots
\label{recreln}\end{equation}
with the initial condition $P_0=1$, supplemented by the
condition that $P_n$ is homogeneous of weight $2n$, where the $n$-th
$x$-derivative of $u$ is assigned weight $n+2$.
The KdV hierarchy is then the set of
differential equations
\begin{equation} 
u_{t_n} = -\partial_x P_{n+1} ~~~~~~~~n=1,2,\ldots.    \label{KdV}
\label{flows}\end{equation}
Here $u$ is regarded as a function of $x$ and the infinite number of 
``times'' $t_n$, $n=1,2,\ldots$. I use a non-standard numbering of 
these times to underscore the fact that I am considering KdV without
using its embedding in KP. The choice $n=1$ is the original KdV equation;
$P_2=-\frac14u_{xx}+\frac32u^2$ (note the conventions used from here on 
in this paper differ from those used in Sec.~\ref{Galas}
where I followed the conventions of \cite{Galas}). Partial
derivatives are denoted by suffices, as usual. The proofs that the recursion
relation (\ref{recreln}) has a unique solution, when supplemented with
the homogeneity condition given, and that the flows (\ref{flows}) 
commute, are by now standard (see, for example, \cite{Olver}). 

The hierarchy can be presented as a set of zero-curvature conditions
\begin{equation}
A_{t_n} = B^{(n)}_x - [A,B^{(n)}] ~~~~~~~n=1,2,\ldots  \label{zcc}
\end{equation}
where 
\begin{eqnarray}
A &=& \pmatrix{0&1\cr 2u+\lambda&0\cr} \\
B^{(n)} &=& \pmatrix{-{\textstyle\frac12}b^{(n)}_x & b^{(n)} \cr
            (2u+\lambda)b^{(n)}-{\textstyle\frac12}b^{(n)}_{xx} &
            {\textstyle\frac12}b^{(n)}_x \cr} \\
b^{(n)} &=& \sum_{r=0}^n P_{n-r}\lambda^r. \label{bndefn}
\end{eqnarray}       
Drinfeld and Sokolov \cite{DS} introduced a generalization of this scheme.
The zero curvature equations (\ref{zcc}) are invariant under gauge
transformations
\begin{eqnarray}
A &\rightarrow& hAh^{-1} + h_xh^{-1}  \\
B^{(n)} &\rightarrow& hB^{(n)}h^{-1} + h_{t_n}h^{-1}, \label{Bgt}
\end{eqnarray}
where $h$ is an $SL(2)$ matrix (which is allowed to be $x,t_n$ 
and even $\lambda$ dependent).
Taking
\begin{equation}
h = \pmatrix{1&0\cr -j&1\cr},    \label{htype}
\end{equation}
where $j$ is dependent on $x,t_n$ but not $\lambda$,
gives a zero curvature system with 
\begin{equation}
A = \pmatrix{j&1\cr 2u-j_x-j^2+\lambda&-j\cr}.
\end{equation}
Choosing $j$ to be a solution of $j_x+j^2=2u$ gives a zero curvature 
formulation for the modified KdV hierarchy. However, I will be more 
interested in a less commonly studied gauge (despite the fact it is 
actually mentioned in \cite{DS}). Take $j$ to be a solution of 
$u=j_x$ in the above; to implement the gauge transformation on the 
matrices $B^{(n)}$ it is necessary to know $j_{t_n}$, and the obvious
choice is
\begin{equation}
j_{t_n}=-P_{n+1},   \label{PKdV}
\end{equation}
which is clearly consistent with the assignment $u=j_x$. This gives
a zero curvature formulation for the potential KdV hierarchy (i.e. the
coupled system of Eqs~(\ref{KdV}) and~(\ref{PKdV}) with the relation
$u=j_x$). I call such a gauge choice ``PKdV gauge''.
Amongst the family of zero curvature equations related by
gauge transformations in the way described above, it turns out that PKdV
gauge has a useful characterization:

\smallskip

\noindent{\bf Lemma.} Up to translation of $j$ by a function of $x$,
independent of $t_n$, PKdV gauge is the unique gauge choice such that
$ B^{(n+1)}-\lambda B^{(n)}$ is independent of $\lambda$.

\smallskip 

\noindent{\bf Proof.} From Eq.~(\ref{bndefn}), $b^{(n+1)}-\lambda
b^{(n)}$ is independent of $\lambda$. After the gauge transformation given
by Eq.~(\ref{Bgt}) with $h$ given by Eq.~(\ref{htype}),
\begin{equation}
B^{(n)} = \pmatrix{-{\textstyle\frac12}b^{(n)}_x+jb^{(n)} & b^{(n)} \cr
            (2u+\lambda)b^{(n)}-{\textstyle\frac12}b^{(n)}_{xx}
            +jb^{(n)}_x-j^2b^{(n)}-j_{t_n}   &
            {\textstyle\frac12}b^{(n)}_x-jb^{(n)} \cr}.
\end{equation}
Thus $ B^{(n+1)}-\lambda B^{(n)}$ is independent of $\lambda$
if and only if $(\lambda b^{(n+1)} - j_{t_{n+1}}) -
\lambda(\lambda b^{(n)} - j_{t_n}) $ is independent of $\lambda$,
which, since the $O(\lambda^0)$ term in $b^{(n+1)}$ is $P_{n+1}$,
is true if and only if $j_{t_n}=-P_{n+1}$. This is the case in PKdV gauge.
It also implies that $u-j_x$ is independent of $t_n$, giving
$u=j_x+f(x)$ where $f(x)$ does not depend on the $t_n$, thus specifying
PKdV gauge, up to translation of $j$ by a function of $x$. $\Box$

\smallskip

It will be useful to have a theorem stating clearly the existence and
characterization of the zero curvature formulation in PKdV gauge.
To this end, let us define $M$ to be the affine space with coordinates
$t_{-1},t_0,t_1,t_2,\dots$, and ${\cal A}$ to be the Lie
algebra of traceless $2\times 2$ matrices whose entries are formal
power series in $\lambda$. I also need the following notion:

\smallskip

\noindent{\bf Definition.} A change of coordinates 
$t_i\rightarrow t'_i$ on $M$  is {\em admissible} if it is given
by 
\begin{equation}
t_i = \sum_{j=0}^{\infty} a_j t'_{i+j} ~~~~~~~~i=-1,0,1,\ldots
\end{equation}
where $a_0=1$, and for $j>0$, the $a_j$ are arbitrary constants,
only a finite number of which are nonzero.
\smallskip

\noindent
I will need the effect of admissible changes of coordinates on
the components of a one-form on $M$. 
If $\alpha=\sum_{n=-1}^{\infty} \alpha_n dt_n
=\sum_{n=-1}^{\infty} \alpha'_n dt'_n$ is a one-form on $M$, then
by a simple substitution,  
\begin{equation}
\alpha'_n=\sum_{m=-1}^n a_{n-m} \alpha_m.
\end{equation}
With these preparations I can now state the following 

\smallskip
\smallskip

\noindent{\bf Theorem.} Let $Z$ be an ${\cal A}$ valued one form on
$M$, i.e.  $Z=\sum_{n=-1}^{\infty} Z_ndt_n$, $Z_n\in{\cal A}$.
Suppose (1) that $Z_0$ has the form $$ Z_0 = \pmatrix{ j&1\cr
\lambda+u-j^2 & -j\cr}, $$ with $j,u$ functions on $M$, (2) that
$Z_{n}-\lambda Z_{n-1}$ is independent of $\lambda$, $n=0,1,\ldots$,
and (3) that
$dZ=Z\wedge Z$. Then, possibly after a sequence of admissible changes of 
coordinates,  (1) $u=j_{t_0}$, (2) $j,u$ solve the potential
KdV hierarchy, with $x$ identified as $t_0$, and (3)
\begin{eqnarray*}
j_{t_{-1}} &=& -1\cr u_{t_{-1}} &=& 0.
\end{eqnarray*}

\smallskip

\noindent{\bf Proof.}
As a preliminary, note that it is straightforward to show that an 
admissible changes of coordinates does not affect assumption (2), 
and since assumption (3) is written in terms of differential forms, 
it evidently is not affected either. Furthermore, assumptions (1) and 
(2) imply
\begin{equation}
Z_{-1}=\pmatrix{0&0\cr1&0\cr},
\end{equation}
and under an admissible change of coordinates $Z_0\rightarrow
Z_0+a_1 Z_{-1}$, so assumption (1) is also not affected: the function
$u$ is just translated. Thus the assumptions in the theorem allow
us to perform admissible changes of coordinates. The KdV flows, however,
are not left invariant under such changes of coordinates; rather
the flows pick up certain linear combinations of the lower flows. This
accounts for the need to allow for an admissible change of coordinates
in the conclusion of the theorem. Fortunately, this cumbersome
but harmless detail will not play any role in the rest of this paper.

The equation $dZ=Z\wedge Z$ is equivalent to
the system
\begin{equation}
\frac{\partial Z_m}{\partial t_n} -
\frac{\partial Z_n}{\partial t_m} + [Z_m,Z_n] =0 ~~~~~m,n=-1,0,1,\ldots
\label{curvs}\end{equation} 
Conclusion (3) in the theorem follows immediately from taking  $m=0$,
$n=-1$. To obtain conclusion (1), let us take $m=0$, $n\ge 1$. Then
\begin{eqnarray}
\frac{\partial Z_0}{\partial t_n}
  & = & \frac{\partial Z_n}{\partial t_0} - [Z_0,Z_n]  \nonumber\\
  & = & \frac{\partial (Z_n-\lambda Z_{n-1})}{\partial t_0} 
      - [Z_0,(Z_n-\lambda Z_{n-1})]  + 
     \lambda \left(\frac{\partial Z_{n-1}}{\partial t_0} - [Z_0,Z_{n-1}]
      \right) \nonumber\\
  & = &  \frac{\partial (Z_n-\lambda Z_{n-1})}{\partial t_0} 
      - [Z_0,(Z_n-\lambda Z_{n-1})]  +  \lambda
     \frac{\partial Z_0}{\partial t_{n-1}}.  \label{messy}
\end{eqnarray} 
Write $\Delta_n=Z_n-\lambda Z_{n-1}$, $n=0,1,2,\ldots$,
which by hypothesis is
independent of $\lambda$. The LHS of Eq.~(\ref{messy}) is independent
of $\lambda$, and the RHS is linear. Setting the coefficient of 
$\lambda$ on the RHS to zero,
\begin{equation}
\frac{\partial Z_0}{\partial t_{n-1}}=
   \left[ \pmatrix{0&0\cr 1&0\cr},\Delta_n \right],
\label{what}\end{equation}
which implies
\begin{eqnarray}
(\Delta_n)_{11} &=& \frac12\frac{\partial}{\partial t_{n-1}}(u-j^2)
   \nonumber\\
(\Delta_n)_{12} &=& -\frac{\partial j}{\partial t_{n-1}} . \label{bitdel}
\end{eqnarray}
The terms independent of $\lambda$ in Eq.~(\ref{messy}) give
\begin{equation}
\frac{\partial Z_0}{\partial t_n}
   =  \frac{\partial \Delta_n}{\partial t_0} -
     \left[\pmatrix{j&1\cr u-j^2&-j\cr},\Delta_n\right] .
\label{28}\end{equation}
The 1,2 entry of this gives
\begin{equation}
     \frac{\partial(\Delta_n)_{12}}{\partial t_0} -
    2(j(\Delta_n)_{12}-(\Delta_n)_{11}) = 0 .
\end{equation}
Substituting from Eq.~(\ref{bitdel}),
\begin{equation}
   \frac{\partial}{\partial t_{n-1}}(u-j_{t_0}) = 0,
\end{equation}
which implies, since it is true for all $n\ge 1$, that $u=j_{t_0}+c$
where $c$ is a function of $t_{-1}$ alone. From the $t_{-1}$ evolution
equations as already established, $c$ must in fact be constant.
Having shown this, $c$ can be removed via an admissible change
of coordinates (with only $a_1$ nonzero), which as explained above
translates $u$ by a constant, thus proving conclusion(1).

To establish conclusion (2) requires looking at the 1,1 and 2,1 
entries of Eq.~(\ref{28}). Using Eq.~(\ref{bitdel}) and the
relation $u=j_{t_0}$ these give
\begin{eqnarray}
\frac{\partial j}{\partial t_n} &=& -(\Delta_n)_{21}
+\frac12\frac{\partial^2}{\partial t_0^2}\frac{\partial j}{\partial t_{n-1}}
-j\frac{\partial}{\partial t_0}\frac{\partial j}{\partial t_{n-1}}
+(j^2-2u)\frac{\partial j}{\partial t_{n-1}}  \label{mad1}\\
\frac{\partial u}{\partial t_n} &=& 
    \frac{\partial(\Delta_n)_{21}}{\partial t_0}
+j\frac{\partial^2}{\partial t_0^2}\frac{\partial j}{\partial t_{n-1}}
-(u+j^2)\frac{\partial}{\partial t_0}\frac{\partial j}{\partial t_{n-1}}
-2ju\frac{\partial j}{\partial t_{n-1}}. \label{mad2}
\end{eqnarray}
Eliminating $(\Delta_n)_{21}$ by differentiating the first of these
equations with respect to $t_0$ and adding to the second, gives,
after some algebra:
\begin{equation}
\frac{\partial u}{\partial t_n} = 
({\textstyle{\frac14}}\partial_{t_0}^3 -  u\partial_{t_0} - \partial_{t_0} u)
\frac{\partial j}{\partial t_{n-1}}.
\label{nearly}\end{equation}
This is almost the required result. It is clear that given
the correct $t_{n-1}$ flow for $j$, viz. $j_{t_{n-1}}=-P_n$, 
then Eq.~(\ref{nearly}) implies the correct $t_n$ evolution for $u$, viz.
$u_{t_n}=-\partial_{t_0}P_{n+1}$. Integrating this with respect to $t_0$ 
gives $j_{t_n}=-P_{n+1}+c_n(t_{-1},t_1,t_2,\ldots)$. To complete an 
inductive step requires showing $c_n$ can be set to zero. This involves
three stages. First, $t_{-1}$ independence follows of $c_n$ follows
immediately from the $t_{-1}$ flow equations for $u,j$ that have already
been established. The second stage is showing there is no $t_1,t_2,\ldots$
dependence. This involves use of Eqs.~(\ref{curvs}) for $m,n\ge 1$, and
is unfortunately intricate, and is therefore relegated to an appendix.
Once it has been established that $c_n$ is constant, the third stage is to 
observe that it can be eliminated by an admissible change of coordinates with 
only $a_{n+1}$ nonzero (from Eq.~(\ref{mad1}), the
addition of a constant to $j_{t_n}$ is equivalent to addition
of a constant to $(\Delta_n)_{21}$, i.e. addition of a constant 
multiple of $Z_{-1}$ to $Z_n$). In this manner  
it emerges that Eq.~(\ref{nearly})
is the crucial ingredient in an inductive step, obtaining the correct 
$t_n$ flow for $j$ from the correct $t_{n-1}$ flow. The induction is 
started from the relation $j_{t_0}=u$, which has already been proven. $\Box$

\section{A Map from the $SL(2,{\bf C})$ Loop Group to Solutions of KdV}
\label{the-constr}
On $M$, define the following one-form valued in the Lie algebra of $G$:
\begin{equation}
\Omega=\sum_{n=-1}^{\infty} \pmatrix{0&1\cr\lambda&0\cr}^{2n+1} dt_n
      =\sum_{n=-1}^{\infty} \pmatrix{0&1\cr\lambda&0\cr}\lambda^ndt_n.
\label{Omega}\end{equation}
Consider on $M$ the linear differential system
\begin{equation}
dU(t)=\Omega U(t),
\label{Uequn}\end{equation}
where $U(t)$ is a $G$ valued function on $M$. Since $d\Omega + 
\Omega\wedge\Omega = 0$ (in fact $d\Omega=\Omega\wedge\Omega=0$) this system
is completely integrable in the sense of Frobenius, and its general solution
is
\begin{equation}
U(t) = 
\exp\left(\sum_{n=-1}^{\infty} \pmatrix{0&1\cr\lambda&0\cr}\lambda^nt_n
\right)U_0,
\label{Usoln}\end{equation}
where $U_0=U(0)$ is an arbitrary element of $G$. Since the
big cell of $G$, where the Birkhoff decomposition can be applied, is an open
dense subset of $G$, it comes as no surprise that the flow on $G$ defined by
(\ref{Usoln}) only leaves the big cell for discrete values of $t$ 
\cite{Wilson}. Let us consider therefore the Birkhoff decomposition of
$U$, in an interval of values of $t$ for which it is defined. Writing
\begin{equation}
U(t) = S^{-1}(t)\cdot Y(t),
\end{equation}
where $Y(t)\in G^+$, $S(t)\in G^-$, and  substituting in
(\ref{Uequn}), gives 
\begin{equation} 
-dS\cdot S^{-1} + dY\cdot Y^{-1} = S\Omega S^{-1}. 
\end{equation}
The first term on the LHS lies in the Lie algebra of $G_-$, whereas
the second term lies in the Lie algebra of $G_+$. Therefore, 
using the obvious notation for the projections of an element in the 
Lie algebra of $G$ to those of $G_+$ and $G_-$,
\begin{eqnarray}
dS\cdot S^{-1}&=& - (S\Omega S^{-1})_- \\
dY\cdot Y^{-1}&=&  (S\Omega S^{-1})_+ . \label{Yflow}
\end{eqnarray}
Write $Z=(S\Omega S^{-1})_+$; Z is an ${\cal A}$ valued one form on
$M$. Using Eq.~(\ref{Yflow}), $dZ=Z\wedge Z$. But using the 
definition of $\Omega$, Eq.~(\ref{Omega}),
\begin{equation}
Z = \sum_{n=-1}^{\infty} Z_n dt_n,~~~~~~~~~
     Z_n=\left( S\pmatrix{0&1\cr\lambda&0\cr}S^{-1}\lambda^n\right)_+.
\label{Zndef}\end{equation}
Now 
\begin{equation}
S = I + \frac1{\lambda}\pmatrix{\alpha&\beta\cr\gamma&-\alpha\cr}
    + O\left(\frac1{\lambda^2}\right)
\label{defS}\end{equation}
for some functions $\alpha,\beta,\gamma$ on $M$. A simple computation
now gives
\begin{equation}
Z_0 = \pmatrix{\beta&1\cr\lambda-2\alpha&-\beta\cr}.
\end{equation}
{}From the definition of the $Z_n$, Eq.~(\ref{Zndef}), it is apparent
that $Z_n-\lambda Z_{n-1}$ is independent of $\lambda$. Thus all
three of the conditions for the theorem of Sec.~\ref{zc} hold, and
it follows that, up to an admissible change of coordinates,
the functions $j=\beta$, $u=\beta^2-2\alpha$ satisfy the potential
KdV hierarchy. In other words, given any
solution of Eq.~(\ref{Uequn}) (a solution of which is specified
by the choice of $U_0$, an element of the loop group),
there is an associated solution of the potential KdV hierarchy.
The replacement $U_0\rightarrow U_0 \cdot g_+$, where
$g_+\in G^+$, has the effect $U\rightarrow U\cdot g_+$ and
$Y\rightarrow Y\cdot g_+$, but $S$ and $Z$ and therefore the associated
KdV solution are left unchanged. Equally easy to see is that the
replacement
\begin{equation}
U_0 \rightarrow  h\cdot U_0,~~~~~~
  h=  \exp\left(\sum_{m=2}^{\infty} \frac{t_{-m}}{\lambda^m}
   \pmatrix{0&1\cr\lambda&0\cr}   \right),
\end{equation}
where $t_{-2},t_{-3},\ldots$ are parameters, causes 
$U\rightarrow h\cdot U$, $S\rightarrow S\cdot h^{-1}$, but $Y$ and $Z$
are left invariant. 
Thus the map is actually from the double coset space $H\backslash G/G^+$ 
to solutions of KdV, where $H$ is the infinite dimensional abelian
subgroup of $G$ generated by the commuting matrices
\begin{equation}
\frac1{\lambda^n}\pmatrix{0&1\cr\lambda&0\cr}~~~~~~~~~n=2,3,\ldots
\end{equation}
in the Lie algebra of $G$. 
The solutions of KdV obtained this way can have singularities at points 
where the Birkhoff decomposition is not possible.

Let us pause to reiterate the important points mentioned in
Sec.~\ref{am}. The loop group element $U_0$ arises as the initial value
data for the simple linear flow Eq.~(\ref{Uequn}), which is the
linear system ``behind'' KdV. There are {\em two} formulae for $Z$,
$Z=(S\Omega S^{-1})_+$ and $Z=dY\cdot Y^{-1}$, meaning that given a 
solution of Eq.~(\ref{Uequn}) the associated KdV solution can be 
constructed either going through the flow on $G^+$ given by 
$Y$, or through the flow on $G^-$ given by $S$. Finally,  in addition
to the $G^+$ action, there is an
$H$ action on $U_0$ that leaves the KdV solution invariant.

One way of thinking about this extra invariance is as follows: I have chosen
to work on the affine space with coordinates $t_{-1},t_0,\dots$, but I could
have decided to work on  the affine space with coordinates 
$\ldots,t_{-2},t_{-1},t_0,t_1,t_2,\ldots$, taking the sum in the definition
of $\Omega$, Eq.~(\ref{Omega}), to be from $-\infty$ to $\infty$ rather than
from $-1$ to $\infty$. The new flows would all leave $Y$ and $Z$ invariant,
but not $S$. It follows that there must be degrees of freedom in $S$ that
do not ``contribute'' to $Z$, that are acted upon by the new flows. This
is indeed the case; the simplest example of such a degree of freedom is
$\gamma$ in Eq.~(\ref{defS}) above. Since these new flows evidently commute,
when working on the affine space with 
coordinates $t_{-1},t_{0},\ldots$, there must be an infinite dimensional
abelian group action on $U_0$ leaving $Z$, but not $S$, invariant;
this is just the $H$ action introduced above. 

For later reference I give a number of formulae. First, I give the
relation of the first few terms in $Y$ and the solution functions $j,u$. 
Writing
\begin{equation}
Y = \sum_{n=0}^{\infty} Y_n\lambda^n,
\end{equation}
where the matrices $Y_n$ are independent of $\lambda$,
\begin{eqnarray}
Z_0 &=& \frac{\partial Y}{\partial t_0} Y^{-1} \nonumber \\
    &=& \left( \frac{\partial Y_0}{\partial t_0} +
               \frac{\partial Y_1}{\partial t_0}\lambda \right)
        \left( Y_0^{-1} - Y_0^{-1}Y_1Y_0^{-1}\lambda\right)
        +O(\lambda^2)   \nonumber \\
    &=& \frac{\partial Y_0}{\partial t_0}Y_0^{-1}
    + Y_0\frac{\partial (Y_0^{-1}Y_1)}{\partial t_0}Y_0^{-1} 
       \lambda + O(\lambda^2). \label{ctoan}
\end{eqnarray}
Writing 
\begin{equation}
Y_0 = \pmatrix{a_1&a_2\cr a_3&a_4\cr}~~~~~~~~~~~~~
Y_0^{-1}Y_1 = \pmatrix{b_1&b_2\cr b_3&-b_1\cr},
\label{Ycpts}\end{equation}
the following relationships emerge (in addition to the determinant 
constraint $a_1a_4-a_2a_3=1$):
\begin{eqnarray}
a_3=a_{1t_0}-ja_1~~~~~~~&~~&~~~~~~~~a_{1t_0t_0}=2ua_1  \nonumber\\
a_4=a_{2t_0}-ja_2~~~~~~~&~~&~~~~~~~~a_{2t_0t_0}=2ua_2  \nonumber\\
    b_{1t_0} &=& -a_1a_2   \label{strucequns}\\
    b_{2t_0} &=& -a_2^2   \nonumber\\
    b_{3t_0} &=& a_1^2.  \nonumber
\end{eqnarray}
Evolutions for components of $Y$ can easily be determined using the
higher components of $Z=dY\cdot Y^{-1}$; in particular, from $Z_1=
(\partial Y/\partial t_1)\cdot Y^{-1}$,
\begin{eqnarray}
a_{1t_1}&=&  {\textstyle{\frac12}}u_{t_0}a_1-ua_{1t_0}\nonumber \\
a_{2t_1}&=&  {\textstyle{\frac12}}u_{t_0}a_2-ua_{2t_0}\nonumber \\ 
b_{1t_1}&=&  a_{1t_0}a_{2t_0}-ua_1a_2 \label{t1els}\\
b_{2t_1}&=&  a_{2t_0}^2-ua_2^2 \nonumber \\
b_{3t_1}&=&  -a_{1t_0}^2+ua_1^2 \nonumber
\end{eqnarray}
Next, consider the problem of to what extent $S$ can be reconstructed
from $Z$. Looking at Eq.~(\ref{Zndef}), and recalling from Sec.~\ref{zc}
that $\Delta_n=Z_n-\lambda Z_{n-1}$, $n=0,1,\ldots$, we
see at once that $\Delta_n$ is the coefficient of $\lambda^{-n}$ in
$S\pmatrix{0&1\cr\lambda&0\cr} S^{-1}$. That is, 
\begin{equation}
S\pmatrix{0&1\cr\lambda&0\cr} S^{-1} =
\pmatrix{0&0\cr\lambda&0\cr} + \sum_{n=0}^{\infty}
    \frac{\Delta_n}{\lambda^n}  =
\sum_{n=-1}^{\infty}\frac{\Delta_n}{\lambda^n} 
\label{SmSinv}\end{equation}
where $\Delta_{-1}\equiv\pmatrix{0&0\cr1&0\cr}$. Finally,
it will be useful to an abbreviation for the matrix appearing in
Eq.~(\ref{Usoln}), so define
\begin{equation}
M =  \exp\left(\sum_{n=-1}^{\infty} \pmatrix{0&1\cr\lambda&0\cr}
     \lambda^nt_n\right)
  = \cosh(z\sqrt{\lambda}) I
    + \frac{\sinh(z\sqrt{\lambda})}{\sqrt{\lambda}}
     \pmatrix{0&1\cr\lambda&0\cr},
\label{Mdef}\end{equation}
where $z=\sum_{n=-1}^{\infty} t_n\lambda^n$, 
so that Eq.~(\ref{Usoln}) now reads $U=MU_0$.

Two more notes are in order before concluding this section. First, I must
mention again that the above construction is modeled on Mulase's similar 
construction for KP \cite{Mulase}; I have followed Mulase's notation 
throughout. Second, the PKdV gauge of the zero curvature formulation 
evidently plays a pivotal role in the construction; I expect though that a
similar formulation for KdV in its standard gauge could be found using a
nonstandard Birkhoff decomposition of $G$ (c.f. the different projections
used in \cite{DS}, Sec.3). 

\section{Symmetry Actions on the Loop Group and Symm\-etries 
of KdV} \label{expl-syms}

In this section I give the symmetry actions on $G$ that give rise 
to the symmetries of KdV listed in Sec.~\ref{symms}. The order
of subsections differs from that of Sec.~\ref{symms} in that I deal with
infinitesimal symmetries first.

\subsection{Infinitesimal symmetries}
\label{inf2}

Consider infinitesimal left multiplications on $U_0$, i.e. 
transformations 
\begin{equation}
U_0 \mapsto \left( I + \epsilon P \right) U_0,
\label{U0inftr}\end{equation}
where $\epsilon$ is an infinitesimal parameter and $P$ is in the Lie
algebra of $G$. If $P$ is independent of $U_0$ this 
defines a map from $G/G^+$ to $G/G^+$,
but it is not clear whether this descends to the double coset space
$H\backslash G/G^+$. Under the transformation (\ref{U0inftr}), 
\begin{eqnarray}
U &\mapsto& M\left( I + \epsilon P \right) U_0
      \nonumber \\
 &=& \left( I + \epsilon MPM^{-1} \right)  U , 
\label{Uinftr}\end{eqnarray}
which gives
\begin{eqnarray}
Y &\mapsto& \left( I + \epsilon (SMPM^{-1}S^{-1})_+ \right) Y \\
S &\mapsto& \left( I - \epsilon (SMPM^{-1}S^{-1})_- \right) S .
\end{eqnarray}
(These are easily computed: an infinitesimal variation $\delta U$ in $U$
gives rise to infinitesimal variations $\delta Y$ in $Y$  and $\delta S$ in
$S$ where $\delta U = - S^{-1} \delta S S^{-1} Y + S \delta Y$, i.e.
$S \delta U Y^{-1} = - \delta S S^{-1} + \delta Y Y^{-1}$. Thus
$\delta S S^{-1} = - (S \delta U Y^{-1})_-$ and
$\delta Y Y^{-1} = (S \delta U Y^{-1})_+$.)  Using now the definition of
$Z_0$  from Eq.~(\ref{Zndef}), a straightforward calculation gives
\begin{equation}
Z_0 \mapsto Z_0 + \epsilon \left[ \pmatrix{0&0\cr1&0\cr},
      {\rm res}(SMPM^{-1}S^{-1}) \right],
\label{Z0ch}\end{equation}
where ${\rm res}(T)$, for $T$ in the Lie algebra of $G$, denotes the 
$O(\lambda^{-1})$ term in $T$.

Consider two special cases. First, take
\begin{equation} 
P = P_n \equiv \lambda^n\pmatrix{0&1\cr\lambda&0\cr} = 
   \pmatrix{0&1\cr\lambda&0\cr}^{2n+1}~~~~~~n=\ldots,-1,0,1,\ldots
\end{equation}
Evidently $MP_nM^{-1}=P_n$. For $n<-1$ these transformations have
no effect on $Z_0$; they are just infinitesimal $H$ transformations.
For $n\ge -1$, using Eq.~(\ref{SmSinv}), 
\begin{eqnarray}
Z_0 &\mapsto& Z_0 +\epsilon \left[\pmatrix{0&0\cr1&0\cr},\Delta_{n+1}
   \right]      \nonumber\\
   & = & Z_0 +\epsilon\frac{\partial Z_0}{\partial t_n},
\end{eqnarray}
where in the last equality I have used Eq.~(\ref{what}). 
Thus the choice
$P=P_n$ generates translations in $t_n$, for $n\ge -1$. 
Since all the matrices $P_n$ commute, these transformation clearly 
descend to the double coset space. I denote
the generator of these transformations on $U_0$ by $l_n$, in accordance 
with the notation of Sec.~\ref{infsys}. The generator of the infinitesimal
transformation $U_0 \mapsto (1+\epsilon P) U_0 $ is defined as the
operator whose action on $U_0$ gives $P U_0$; so $l_n$ is just 
multiplication by $P_n$. Note that here $l_n$ is defined in
a wider sense than in Sec.~\ref{infsys}; there $l_n$ denoted the generator 
of translations acting on the solutions of KdV, whereas here
it denotes the generator of the transformations acting on the loop group 
that induce translations on solutions of KdV. In particular $l_n$ here
is not zero for negative $n$.

Now let us take
\begin{equation}
P=Q_n\equiv \lambda^n \left( \lambda \frac{dU_0}{d\lambda} U_0^{-1}
     +{\textstyle{\frac14}}\pmatrix{1&0\cr0&-1} \right)~~~~~~~
    n=\ldots,-1,0,1,\ldots
\label{Qndef}\end{equation}
These look like field dependent transformations, since $U_0$ appears
in $P$! But in fact this is just the choice of $P$ for the field
independent infinitesimal transformation which is a linear combination
of a reparametrization of $U_0$ and a multiplication by a fixed 
algebra element:
\begin{equation}
U_0(\lambda) \mapsto U_0(\lambda(1+\epsilon \lambda^n))+
   {\textstyle{\frac14}} \epsilon \lambda^n \pmatrix{1&0\cr0&-1\cr} 
   U_0(\lambda).
\end{equation}
The associated generators are
\begin{equation}  
m_n = \lambda^n\left( \lambda \frac{d}{d\lambda} + {\textstyle{\frac14}}
       \pmatrix{1&0\cr0&-1\cr} \right)
{}    ~~~~~n=\ldots,-1,0,1,\ldots
\label{mndef}\end{equation}
In Sec.~\ref{TGS} I will give the finite action which these generate;
below I will explain why I am  looking at these particular 
transformations. But first let me give the action on the KdV solution.
To use Eq.~(\ref{Z0ch}) it is necessary to compute $SMQ_nM^{-1}S^{-1}$ 
with $Q_n$ given by Eq.~(\ref{Qndef}). Substituting
$U_0=M^{-1}U=M^{-1}S^{-1}Y$ in $Q_n$ gives, after some algebra,
\begin{eqnarray}
SMQ_nM^{-1}S^{-1}& = &\lambda^{n+1}(Y'Y^{-1} - S'S^{-1})
        \nonumber\\ &+& \lambda^n S
   \left( {\textstyle{\frac14}}M\pmatrix{1&0\cr0&-1\cr}M^{-1}
     -\lambda M'M^{-1}\right)S^{-1} ,
\end{eqnarray}
where a prime denotes differentiation with 
respect to $\lambda$. Using Eq.~(\ref{Mdef}) 
\begin{equation}
{\textstyle{\frac14}}M\pmatrix{1&0\cr0&-1\cr}M^{-1}
     -\lambda M'M^{-1}
={\textstyle{\frac14}}\pmatrix{1&0\cr0&-1\cr} -
  \sum_{m=-1}^{\infty}(m+{\textstyle{\frac12}})t_m\lambda^m
  \pmatrix{0&1\cr\lambda&0\cr}.
\end{equation}
The last two equations and Eq.~(\ref{SmSinv}) give the final result
\begin{eqnarray}
SMQ_nM^{-1}S^{-1}& = &\lambda^{n+1}(Y'Y^{-1} - S'S^{-1}) + 
  \frac{\lambda^n}4S\pmatrix{1&0\cr0&-1}S^{-1}
        \nonumber\\ &-& \lambda^n 
\left(\sum_{m=-1}^{\infty}(m+{\textstyle{\frac12}})t_m\lambda^m \right)
\left(\sum_{p=-1}^{\infty}\Delta_p \lambda^{-p} \right).
\label{formula}\end{eqnarray}
The residue of this must be computed. Of the four 
terms, the simplest to discuss is the last, despite its apparent
complexity. For all $n$, it contributes an infinite series of terms
to the residue
\begin{equation}
-\sum_{m={\rm max}(-1,-2-n)}^{\infty} (m+{\textstyle{\frac12}})t_m
      \Delta_{m+n+1}.
\end{equation}
Because this term is given purely in terms of $j$ and $u$, via the
$\Delta_p$'s, it descends to the double coset space. For the other terms
this is not true in general. The first term in (\ref{formula}) will
contribute to the residue if $n\le -2$, since $Y'Y^{-1}$ is a power series
in $\lambda$; the terms it contributes descend to $H\backslash G$, but not
necessarily to the double coset space. The second term will contribute to
the residue if $n \ge 0$ and third term if $n\ge -1$; their contributions
descend to $G/G^+$ but not necessarily to the double coset space. 
So it is not clear that there is ever a genuine symmetry of
KdV here! In practice, however, it turns out that for $n=-1,0$ only
$H$-invariant degrees of freedom from $S$ contribute to the variation
of $Z_0$. A little further calculation gives the results
\begin{eqnarray}
n=-1 ~~~~~~~Z_0 &\mapsto& Z_0 +\epsilon
    \left(  \pmatrix{0&0\cr{1\over 2}&0\cr} -
            \sum_{m=0}^{\infty}(m+{\textstyle{1\over 2}})t_m
            {{\partial Z_0}\over{\partial t_{m-1}}}\right)  \nonumber \\
 j &\mapsto& j - \epsilon\sum_{m=0}^{\infty}
       (m+{\textstyle{1\over 2}})t_m
       {{\partial j}\over{\partial t_{m-1}}} \label{infGal}\\
 u &\mapsto& u + {\epsilon\over2}
       - \epsilon\sum_{m=1}^{\infty}
       (m+{\textstyle{1\over 2}})t_m
       {{\partial u}\over{\partial t_{m-1}}} \nonumber \\
n=0 ~~~~~~~Z_0 &\mapsto& Z_0 +\epsilon
    \left(  \pmatrix{-\beta/2&0\cr2\alpha&\beta/2\cr} -
            \sum_{m=-1}^{\infty}(m+{\textstyle{1\over 2}})t_m
            {{\partial Z_0}\over{\partial t_{m}}}\right)  \nonumber \\
 j &\mapsto& j - {{\epsilon j}\over 2}
       -\epsilon\sum_{m=-1}^{\infty}
       (m+{\textstyle{1\over 2}})t_m
       {{\partial j}\over{\partial t_{m}}} \\
 u &\mapsto& u  - \epsilon u
       - \epsilon\sum_{m=0}^{\infty}
       (m+{\textstyle{1\over 2}})t_m
       {{\partial u}\over{\partial t_{m}}} .\nonumber
\end{eqnarray} 
The $n=-1$ and $n=0$ cases correspond to Galilean transformations
and scaling respectively. For $n=1$, a detailed calculation gives
\begin{eqnarray}
u&\mapsto& u+\epsilon\left(
{\textstyle{\frac12}}ju_{t_0}+2u^2-{\textstyle{\frac12}}u_{t_0t_0}-
\sum_{m=-1}^{\infty}(m+{\textstyle{\frac12}})t_m
\frac{\partial u}{\partial t_{m+1}}
\right)\\
&=& u+\epsilon\left(
 {\textstyle{\frac12}}(j+t_{-1})u_{t_0}+(2u^2-
 {\textstyle{\frac12}}u_{t_0t_0})
  -{\textstyle{\frac12}}t_0u_{t_1}-
  {\textstyle{\frac32}}t_1u_{t_2}-\ldots\right).
\nonumber\end{eqnarray}
In the second line here I have emphasized the cancellation between terms
that ensures that even though $j$ appears in this transformation law,
$u$ develops no $t_{-1}$ dependence. The $n=1$
transformation law for $j$ involves the non $H$-invariant
quantity $\gamma$ defined in Eq.~(\ref{defS}); and
for higher $n$ other non $H$-invariant terms appear in the 
transformation laws for both $u$ and $j$. This is perfectly in accord
with existing knowledge about the higher symmetries in the hierarchy 
of Galilean and scaling symmetries; each time the recursion operator is 
applied to obtain a higher symmetry, a new integration constant appears.
It is straightforward to check that while the symmetries $m_n$, $n\ge 1$
cannot be considered to descend to the double coset space 
$H\backslash G/G^+$, they do descend to spaces of the form 
$H_n\backslash G/G^+$, where, for every $n$, $H_n$ is a subgroup of $H$
of finite codimension.

Turning now briefly to the $n=-2,-3,\ldots$ cases, none of these symmetries
descend to $G/G^+$. Since it is straightforward to compute
using Eq.~(\ref{formula}), I give the explicit form of the $n=-2$ symmetry:
\begin{eqnarray}
u&\rightarrow& u+ \epsilon\left(2b_1(a_1a_2)_{t_0}-b_2(a_1^2)_{t_0}+
   b_3(a_2^2)_{t_0} - \sum_{m=2}^{\infty}(m+\frac12)t_m\frac{\partial u}
   {\partial t_{m-2}}\right)   \nonumber \\
j&\rightarrow& j+ \epsilon\left(2b_1a_1a_2-b_2a_1^2+
   b_3a_2^2 - \sum_{m=1}^{\infty}(m+\frac12)t_m\frac{\partial j}
   {\partial t_{m-2}}\right),
\end{eqnarray}
where here I am using the notation set up in  Eq.~(\ref{Ycpts}) 
(c.f. \cite{GH1}).

As a final comment on the $m_n$ symmetries, I note 
that the operators defined on the loop group, viz.
\begin{equation}
l_n=\lambda^n\pmatrix{0&1\cr\lambda&0\cr}~~~~~~~
m_n=\lambda^n\left( \lambda \frac{d}{d\lambda} + {\textstyle{\frac14}}
       \pmatrix{1&0\cr0&-1\cr} \right)
\end{equation}
satisfy the commutation relations of Eq.~(\ref{infsymsalg}), where now
all indices are allowed to run over all the integers. They are
related by recursion formulae $l_{n+1}=\lambda l_n$,
and $m_{n+1}=\lambda m_n$; $\lambda$ is playing the role of
the recursion operator. 

To conclude this section, I need to answer the question of why I
have not looked at other infinitesimal symmetries on $G$.
It is easy to check that a generic choice 
of field independent infinitesimal transformation of the form  of
Eq.~(\ref{U0inftr}) simply does not descend to the double coset space. These 
are the messy symmetries discussed in \cite{GH2}. On the other hand,
it is currently not at all clear that I have identified 
the full set of symmetries that do descend to the double coset space. This
is the problem in understanding the (local) geometry of the double coset
space $H\backslash G/G^+$; particularly it is of interest to identify a 
basis for vector fields on this space.

\subsection{Translations, Galilean and scaling symmetries}
\label{TGS}

Consider now the finite transformations generated by the infinitesimal
transformations of Sec.~\ref{inf2}.  Finite translations correspond
to transformations
\begin{equation}
U_0\mapsto\exp\left(\theta \lambda^n\pmatrix{0&1\cr\lambda&0\cr}
       \right)U_0 
\end{equation}
on $G$. This is evidently equivalent to a translation of $t_n$ by
$\theta$, for $n\ge -1$; for $n<-1$ it does not act on KdV solutions.
To exponentiate the symmetries in the hierarchy of Galilean and scaling
symmetries,  observe that
\begin{equation}
m_n=\lambda^{n+1}
     \pmatrix{\lambda^{-{1\over 4}}&0\cr0&\lambda^{{1\over 4}}\cr}
     {d\over {d\lambda}} 
     \pmatrix{\lambda^{{1\over 4}}&0\cr0&\lambda^{-{1\over 4}}\cr},
\end{equation}
so a finite transformation generated by this takes the form
\begin{equation}
U_0(\lambda) \mapsto \pmatrix{(1+\theta\lambda^n)^{1\over 4}&0\cr
                         0&(1+\theta\lambda^n)^{-{1\over 4}}\cr}
     U_0\left(\lambda(1+\theta\lambda^n)\right).
\label{hid}\end{equation}
The quantity on the RHS here will not be in $G$ for arbitrary $U_0$
and $\theta$, but I assume the necessary restrictions are imposed so that
it is. When do these transformations descend to $H\backslash G/G^{+}$?
To descend to the coset space $H\backslash G$,
the effect of an $H$ transformation on $U_0$ followed by a transformation
of the above type must be 
equivalent to the effect of first applying the
transformation of the above type followed by a (possibly different) 
$H$ transformation. That is, for all choices of $t_{-2},t_{-3},\ldots$
it must be possible to find $s_{-2},s_{-3},\ldots$ such that
\begin{eqnarray}
&{\displaystyle \pmatrix{(1+\theta\lambda^n)^{1\over 4}&0\cr
                         0&(1+\theta\lambda^n)^{-{1\over 4}}\cr}
\exp\left( \sum_{m=2}^{\infty} {{t_{-m}}\over{(\lambda
     (1+\theta\lambda^n))^m}}\pmatrix{0&1\cr\lambda(1+\theta\lambda^n)
     &0}\right)}&
\nonumber \\
&{\displaystyle =\exp\left( \sum_{m=2}^{\infty} {{s_{-m}}\over{\lambda^m
     }}\pmatrix{0&1\cr\lambda&0}\right)
\pmatrix{(1+\theta\lambda^n)^{1\over 4}&0\cr
                         0&(1+\theta\lambda^n)^{-{1\over 4}}\cr}}&
\end{eqnarray}
Multiplying this out, various cancellations take place, and it turns out
that this is equivalent to the simple requirement
\begin{equation}
\sum_{m=2}^{\infty} {{s_{-m}}\over{\lambda^m}} =
\sum_{m=2}^{\infty} {{t_{-m}}\over{\lambda^m}}
   (1+\theta\lambda^n)^{-m+{1\over 2}}
\end{equation}
For $n\le0$ this can be  satisfied, taking $\vert\theta\vert < 1$
and expanding in negative powers of $\lambda$ for $\vert\lambda\vert>
\vert\theta\vert$. However, it is obvious that symmetries of the form
of Eq.~(\ref{hid}) only descend to $G/G^+$ if for all $g_+(\lambda)\in G^+$,
$g_{+}(\lambda(1+\theta\lambda^n))\in G^+$. This requires $n\ge -1$.
Thus, as expected from the analysis in the infinitesimal case, the only
symmetries that descend to the double coset space are the cases $n=-1,0$,
the cases of Galilean and scaling symmetries respectively. In particular,
these results show that {\em the so-called hidden symmetries of KdV cannot
be exponentiated}.

It remains to give formulae for finite Galilean and scaling transformations.
For $n=0$ (scaling transformations)
\begin{equation}
U_0(\lambda)\mapsto
\pmatrix{p^{\frac12}&0\cr0&p^{-\frac12}\cr}U_0(p^2\lambda),
\end{equation} 
where $p=(1+\theta)^{\frac12}$. A simple calculation gives
\begin{equation}
U(\lambda,t)\mapsto
\pmatrix{p^{\frac12}&0\cr0&p^{-\frac12}\cr}
U(p^2\lambda,s),
\end{equation} 
where 
\begin{equation}
s_m = \frac1{p^{2m+1}} t_m ~~~~~~~~~~m=-1,0,\ldots 
\end{equation} 
In turn this gives 
\begin{eqnarray}
S(\lambda,t) &\mapsto& 
\pmatrix{p^{1\over 2}&0\cr0&p^{-{1\over 2}}\cr}
S\left(p^2\lambda,s\right) 
\pmatrix{p^{-{1\over 2}}&0\cr0&p^{{1\over 2}}\cr} \\
Y(\lambda,t) &\mapsto& 
\pmatrix{p^{1\over 2}&0\cr0&p^{-{1\over 2}}\cr}
Y\left(p^2\lambda,s\right), \nonumber
\end{eqnarray}
where I have assumed that $p$ is such that the matrices on the RHS of
these expressions are genuinely in $G^-,G^+$ respectively. Finally from
the $O(\lambda^{-1})$ term of the $S$ transformation law, emerge the
standard scaling transformations
\begin{equation}
j(t) \mapsto \frac1p j(s),~~~~~~~~~~~~
u(t) \mapsto \frac1{p^2} u(s). 
\end{equation}

For $n=-1$ (Galilean transformations) things are marginally more
difficult. The starting point is
\begin{equation}
U_0(\lambda)\mapsto
\pmatrix{(1+\theta/\lambda)^{\frac14}&0\cr0&
      (1+\theta/\lambda)^{-\frac14}\cr}U_0(\lambda+\theta).
\end{equation} 
This gives, after some effort:
\begin{eqnarray}
U(\lambda,t) &\mapsto& \pmatrix{(1+\theta/\lambda)^{1\over 4}&0\cr
                         0&(1+\theta/\lambda)^{-{1\over 4}}\cr}
    \exp\left(\sum_{m=2}^{\infty}\frac{s_{-m}}{(\lambda+\theta)^m}
     \pmatrix{0&1\cr\lambda+\theta&0\cr}    \right) \nonumber\\
  &&     U\left((\lambda+\theta),s\right) 
\end{eqnarray}
where the ``times'' $\ldots,s_{-2},s_{-1},s_0,\ldots$ are related to the 
``times'' $t_{-1},t_0,\ldots$ by
\begin{equation}
\sum_{m=-1}^{\infty} t_m\lambda^{m+\frac12} =
\sum_{m=-\infty}^{\infty} s_m\left(\lambda
   +\theta\right)^{m+\frac12},
\end{equation}
This in turn gives
\begin{eqnarray}
S(\lambda,t) &\mapsto& S\left(\lambda+\theta,s\right) 
\exp\left(-\sum_{m=2}^{\infty}\frac{s_{-m}}{(\lambda+\theta)^m}
     \pmatrix{0&1\cr\lambda+\theta&0\cr}    \right)   \nonumber\\
&&\pmatrix{(1+\theta/\lambda)^{-{1\over 4}}&0\cr0&
  (1+\theta/\lambda)^{{1\over 4}}\cr}    \\
Y(\lambda,t) &\mapsto&   
Y\left(\lambda+\theta,s\right), \nonumber
\end{eqnarray}
where, again, certain assumptions on $\theta$ have been made;
finally, the Galilean invariance formulae emerge:
\begin{eqnarray}
j(t) &\mapsto&  j(s)  ~~~~~~~~~~
u(t) ~~\mapsto~~ u(s) + {\textstyle{\frac{\theta}2}} \label{fintr}\\
s_m &=& \sum_{n=m}^{\infty} \pmatrix{n+\frac12\cr n-m\cr} (-\theta)^{n-m}
      t_n ~~~~~~~~~~m=-1,0,\ldots \nonumber
\end{eqnarray}
The reader may wish to try deriving these from both the transformation 
law for $Y$  and that for $S$; the latter is much simpler to use.
The expressions in Eq.~(\ref{fintr}) for finite Galilean transformations 
are consistent with the infinitesimal transformation
law Eq.~(\ref{infGal}), and indeed it can be checked directly
that Eq.~(\ref{fintr}) does give a symmetry of the whole KdV
hierarchy. For future reference, I call this Galilean transformation
$A(\theta)$. The range of values of $\theta$ for which the
transformation $A(\theta)$ is defined on $G$ is limited, and is different
for different elements of $G$. For the action of $A(\theta)$ on $u,j$
there is also a limitation, of convergence of the series
in Eq.~(\ref{fintr}), but these limitations do not correspond. This is 
just the first symptom to appear of the fact that the class of
KdV solutions being studied in this paper is
limited. The restrictions on applicability of $A(\theta)$ will be 
important in the next section.

\subsection{Wahlquist-Estabrook B\"acklund Transformations}
\label{WEBT}

Let $B:G\rightarrow G$ be the involution on $G$ defined by
\begin{equation} 
U_0 \mapsto \pmatrix{0&1\cr\lambda&0\cr} U_0
                \pmatrix{0&\lambda^{-1}\cr1&0\cr}
\end{equation}
This descends to an involution on the coset space $H\backslash G$, since
$U_0$ is multiplied on the left by a matrix that commutes with all elements
of $H$. It does not descend to $H\backslash G/G^+$ though, because
\begin{equation}
U_0 \pmatrix{a&b\cr c&d\cr} \pmatrix{0&\lambda^{-1}\cr 1&0\cr}
= U_0 \pmatrix{0&\lambda^{-1}\cr 1&0\cr}
      \pmatrix{d&c\lambda^{-1}\cr\lambda b&a\cr},
\end{equation}
so right multiplication of $U_0$ by an element of $G^+$
followed by application of $B$ is not equivalent to application of
$B$ followed by right multiplication by a (possibly different) element
of $G^+$. The above identity shows, however, that $B$ {\em does}
descend to the double coset space $H\backslash G/J$, where $J$ is
the subgroup of $G^+$ consisting of matrices $\pmatrix{a&b\cr
c&d\cr}\in G^+$  such that $c$ has no constant term in its power series
expansion in $\lambda$. $J$ is a codimension 1 subgroup of $G^+$, so
$B$ should give rise to a map that generates a one parameter family
of solutions of KdV from a given one. 

Defining $G^{+(1)}$ to be the subgroup of elements of $G^+$ that reduce
to the identity at $\lambda=0$,  $G^{+(1)}\subset J\subset G^+$,
and $G^{+(1)}$ is normal in $G^+$.  $G^+/G^{+(1)}$ is naturally identified
as the subgroup of constant elements in $G$. It follows that if 
I define the map ${\cal O}_0$, mapping $G$ to the space
of subsets of $G$, by
\begin{equation}
U_0 \mapsto \pmatrix{0&1\cr\lambda&0\cr} U_0
 \pmatrix{p&q\cr r&s \cr}  \pmatrix{0&\lambda^{-1}\cr1&0\cr},
\label{fibt}\end{equation} 
where $\pmatrix{p&q\cr r&s \cr}$ is an arbitrary $SL(2,{\bf C})$ matrix,
then this map {\em will} descend to the double coset space 
$H\backslash G/G^+$. It seems at first glance that this map should give
a  three parameter family of solutions from a single one. But the 
full information in the matrix  $\pmatrix{p&q\cr r&s \cr}$ is not actually
relevant; it can be multiplied on the right by any upper triangular matrix 
without effect. The map (\ref{fibt}) gives
\begin{eqnarray}
U &\mapsto& \pmatrix{0&1\cr\lambda&0\cr} U
 \pmatrix{p&q\cr r&s \cr}  \pmatrix{0&\lambda^{-1}\cr1&0\cr} \nonumber\\
 &=& \pmatrix{0&1\cr\lambda&0\cr} S^{-1}\cdot Y
 \pmatrix{p&q\cr r&s \cr}  \pmatrix{0&\lambda^{-1}\cr1&0\cr} \nonumber\\
 &=& \pmatrix{0&1\cr\lambda&0\cr} S^{-1}
 \pmatrix{\frac{\beta}{\lambda}&\frac1{\lambda}\cr
         1+\frac{T\beta}{\lambda}&\frac{T}{\lambda}\cr} \cdot \\
&& ~~~~~~~ \pmatrix{-T&1\cr\lambda+T\beta&-\beta\cr} Y
 \pmatrix{p&q\cr r&s \cr}  \pmatrix{0&\lambda^{-1}\cr1&0\cr} \nonumber
\end{eqnarray}
where in the last line I have inserted a certain matrix and its
inverse, and 
\begin{equation}
T = \frac{pa_3+ra_4}{pa_1+ra_2}=\frac{\tau_{t_0}}{\tau}-j,~~~~~~~~
      \tau=pa_1+ra_2.
\label{taudef}\end{equation}
(Throughout, I use the notation of Sec.~\ref{the-constr} for the components 
of $S$ and $Y$ needed. In particular $\beta$ is defined in 
Eq.~(\ref{defS}), and $a_1,\ldots,a_4$ in Eq.~(\ref{Ycpts}).)
The reason for the insertion in the last equation is that --- as can be
checked by a straightforward calculation --- the RHS is now written in
Birkhoff factorized form, that is
\begin{eqnarray}
S &\mapsto& 
 \pmatrix{\frac{T}{\lambda}&-\frac1{\lambda}\cr
         -1-\frac{T\beta}{\lambda}&\frac{\beta}{\lambda}\cr}      
 S \pmatrix{0&-1\cr-\lambda&0\cr}         \\
Y &\mapsto&  
 \pmatrix{-T&1\cr\lambda+T\beta&-\beta\cr} Y
 \pmatrix{p&q\cr r&s \cr}  \pmatrix{0&\lambda^{-1}\cr1&0\cr}. \nonumber
\end{eqnarray}
Computing the $O(\lambda^{-1})$ terms of the new $S$ gives the
transformation
\begin{eqnarray}
u &\mapsto& \left(\frac{\tau_{t_0}}{\tau}\right)^2 -u 
       = u - \left(\frac{\tau_{t_0}}{\tau}\right)_{t_0} \\
         j &\mapsto& j -\frac{\tau_{t_0}}{\tau}, \nonumber
\end{eqnarray}
where in the first of these equations I have used the result
$\tau_{t_0t_0}=2u\tau$ which is evident from the definition of $\tau$,
Eq.~(\ref{taudef}), and Eq.~(\ref{strucequns}). Using Eq.~(\ref{t1els}), it
is easy to
compute $\tau_{t_1}=\frac12u_{t_0}\tau-u\tau_{t_0}=\frac14(\tau_{t_0t_0t_0}
-3\tau_{t_0t_0}\tau_{t_0}/\tau)$. (Compare Eqs.~(\ref{f1})-(\ref{f2}).)

Having defined and discussed the involution $B:G\rightarrow G$, and
the map ${\cal O}_0$ from $G$ to the space subsets of $G$, the
extension of $B$ which descends to $H\backslash G/G^+$, I now define
one parameter generalizations of these maps, $B(\theta):G\rightarrow G$,
and ${\cal O}_{\theta}$, mapping $G$ to the space of subsets of $G$, via
\begin{eqnarray}
B(\theta)&=&A(-\theta)~B~A(\theta)   \\
B(\theta)U_0 &=& \sqrt{\frac{\lambda-\theta}{\lambda}}
   \pmatrix{0&1 \cr\lambda &0\cr}
   U_0 \pmatrix{0&(\lambda-\theta)^{-1}\cr1&0\cr}    \nonumber \\
{\cal O}_{\theta} &=& A(-\theta)~{\cal O}_0~A(\theta)  \label{oth}\\
{\cal O}_{\theta}U_0 &=& \sqrt{\frac{\lambda-\theta}{\lambda}}  
   \pmatrix{0&1\cr\lambda &0\cr}
   U_0 \pmatrix{p&q\cr r&s \cr} 
   \pmatrix{0&(\lambda-\theta)^{-1}\cr1&0\cr}.    \nonumber 
\end{eqnarray}
Almost everything I have said about $B$ and ${\cal O}_0$ is true of 
$B(\theta)$ and ${\cal O}_{\theta}$, in particular $B(\theta)$ is an
involution, which does not descend to $H\backslash G/G^+$, but does
descend to $H\backslash G/J_{\theta}$, where $J_\theta$ is a codimension
1 subgroup of $G^+$, consisting of the matrices $\pmatrix{a&b\cr c&d\cr}$
in $G^+$ such that $c$ has a zero at $\lambda=\theta$. ${\cal O}_\theta$
is an extension of $B(\theta)$ that does descend to  $H\backslash G/G^+$.
There is one significant difference, however, between  ${\cal O}_0$ and
${\cal O}_{\theta}$ for $\theta\not=0$: this is that because $A(\theta)$
is not defined on all of $G$, nor is $B(\theta)$. I will not study in
detail the domains of the maps ${\cal O}_\theta$, but 
for $\vert\theta\vert <1$ they are large enough for the maps to be 
interesting.

Before I compute the effect of ${\cal O}_\theta$ on $u,j$, let us discuss
some algebraic properties of these maps. First, a simple computation 
shows 
\begin{equation}
B(\theta_1)B(\theta_2) U_0 = U_0 \pmatrix{
\sqrt{\frac{\lambda-\theta_1}{\lambda-\theta_2}} & 0 \cr
0 &  \sqrt{\frac{\lambda-\theta_2}{\lambda-\theta_1}}  \cr}.
\label{BB}\end{equation}
In particular, $B(\theta_1),B(\theta_2)$ {\em do not}
commute, for $\theta_1\not=\theta_2$. On the other hand, 
\begin{eqnarray}
{\cal O}_{\theta_1}{\cal O}_{\theta_2} U_0 &=& 
\sqrt{(\lambda-\theta_1)(\lambda-\theta_2)} U_0 \\ &&
\pmatrix{p_2&q_2\cr r_2&s_2\cr}\pmatrix{0&(\lambda-\theta_2)^{-1}\cr1&0\cr}
\pmatrix{p_1&q_1\cr r_1&s_1\cr}\pmatrix{0&(\lambda-\theta_1)^{-1}\cr1&0\cr},
\nonumber\end{eqnarray}
where $\pmatrix{p_1&q_1\cr r_1&s_1\cr},\pmatrix{p_2&q_2\cr r_2&s_2\cr}$ are
arbitrary $SL(2,{\bf C})$ matrices. Using the identity
\begin{eqnarray}
&\pmatrix{p_2&q_2\cr r_2&s_2\cr}\pmatrix{0&(\lambda-\theta_2)^{-1}\cr1&0\cr}
\pmatrix{p_1&q_1\cr r_1&s_1\cr}\pmatrix{0&(\lambda-\theta_1)^{-1}\cr1&0\cr}
=& \label{mxid}\\
&\pmatrix{p_2-zq_2&q_2\cr r_2-zs_2&s_2\cr}
\pmatrix{0&(\lambda-\theta_1)^{-1}\cr1&0\cr}
\pmatrix{p_1&q_1\cr r_1&s_1\cr}
\pmatrix{0&(\lambda-\theta_2)^{-1}\cr1&0\cr}
\pmatrix{1&0\cr z&1\cr},                      \nonumber
\end{eqnarray}
where $z=(\theta_2-\theta_1)s_1/r_1$ it emerges
that, when interpreted as maps on the coset space $G/G^+$,
${\cal O}_{\theta_1},{\cal O}_{\theta_2}$ {\em do} commute
\footnote{The identity (\ref{mxid}) is only good for $r_1\not=0$. For
$r_1=0$ use
\begin{eqnarray}
&\pmatrix{p_2&q_2\cr r_2&s_2\cr}\pmatrix{0&(\lambda-\theta_2)^{-1}\cr1&0\cr}
\pmatrix{p_1&q_1\cr 0&s_1\cr}\pmatrix{0&(\lambda-\theta_1)^{-1}\cr1&0\cr}
=& \nonumber\\
&i\pmatrix{q_2&p_2\cr s_2&r_2\cr} 
\pmatrix{0&(\lambda-\theta_1)^{-1}\cr1&0\cr}
\pmatrix{0&(\lambda-\theta_2)^{-1}\cr1&0\cr}
(-i)\pmatrix{q_1(\lambda-\theta_1)&p_1\cr s_1&0\cr}.
&\nonumber \end{eqnarray}
}.

The commutativity property of the ${\cal O}_{\theta}$ operators
(on $G/G^+$) is the only one of the properties of Eq.~(\ref{alg-relns}) 
that there is any difficulty establishing; the second property  
follows from the involutiveness of $B(\theta)$. It remains
to check that the operator ${\cal O}_{\theta}$ does indeed act as a 
Backlund transformation; this is a simple generalization for the 
calculation for ${\cal O}_0$. Assuming $Y$ is analytic in a neighborhood
of $\lambda=\theta$, 
\begin{equation}
Y=\sum_{n=1}^{\infty} Y_n^\theta (\lambda-\theta)^n,
\label{Yth1}\end{equation}
where the matrices $Y_n^\theta$ are independent of $\lambda$. Writing
\begin{equation}
Y_0^\theta = \pmatrix{ a_1^\theta & a_2^\theta  \cr
                       a_3^\theta & a_4^\theta  \cr}
\label{Yth2}\end{equation}
gives, from Eq.~(\ref{ctoan}), the structure relations
\begin{eqnarray}
a^\theta_3=a^\theta_{1t_0}-ja^\theta_1~~~~~~~&~~&~~~~~~~~
a^\theta_{1t_0t_0}=(2u+\theta)a^\theta_1  \nonumber\\
a^\theta_4=a^\theta_{2t_0}-ja^\theta_2~~~~~~~&~~&~~~~~~~~
a^\theta_{2t_0t_0}=(2u+\theta)a^\theta_2.
\end{eqnarray}
Starting from the transformation Eq.(\ref{oth}),
\begin{eqnarray}
U &\mapsto& \sqrt{\frac{\lambda-\theta}{\lambda}}
   \pmatrix{0&1\cr\lambda&0\cr} U
 \pmatrix{p&q\cr r&s \cr}  \pmatrix{0&(\lambda-\theta)^{-1}\cr1&0\cr} 
      \nonumber\\
 &=& \sqrt{\frac{\lambda-\theta}{\lambda}}    
 \pmatrix{0&1\cr\lambda&0\cr} S^{-1}
 \pmatrix{\frac{\beta}{\lambda-\theta}&\frac1{\lambda-\theta}\cr
   1+\frac{T^\theta \beta}{\lambda-\theta}&\frac{T^\theta }
             {\lambda-\theta}\cr} \cdot \\
&& ~~~~~~~ \pmatrix{-T^\theta &1\cr\lambda-\theta+T^\theta \beta&-\beta\cr}Y
 \pmatrix{p&q\cr r&s \cr}  \pmatrix{0&(\lambda-\theta)^{-1}\cr1&0\cr} 
  \nonumber\end{eqnarray}
where in the last line I have indicated the Birkhoff factorization
(assuming $\vert\theta\vert < 1$), and 
\begin{equation}
T^\theta = \frac{pa^\theta_3+ra^\theta_4}{pa^\theta_1+ra^\theta_2}=
   \frac{\tau^\theta_{t_0}}{\tau^\theta}-j,~~~~~~~~
      \tau^\theta=pa^\theta_1+ra^\theta_2.
\end{equation}
Computing the $O(\lambda^{-1})$ terms in the transformed $S$ gives
the B\"acklund transformation
\begin{equation}
j \mapsto j - \frac{\tau^\theta_{t_0}}{\tau^\theta}~~~~~~~~~~
u \mapsto u - \left(\frac{\tau^\theta_{t_0}}{\tau^\theta}\right)_{t_0},
\end{equation}
where $\tau^\theta$ satisfies 
$\tau^\theta_{t_0t_0}=(2u+\theta)\tau^\theta$, and (after a little further
calculation)
\begin{equation}
\frac{\tau^\theta_{t_1}}{\tau^\theta_{t_0}} =  
\frac{3\theta}2 + \frac14\left(\frac{\tau^\theta_{t_0t_0t_0}}
    {\tau^\theta_{t_0}}-\frac{3\tau^\theta_{t_0t_0}}{\tau^\theta}
\right).
\end{equation}

It is worth reiterating the algebraic framework that has been hereby 
described for B\"acklund transformations. There is a map between 
cosets of a certain group and solutions of  KdV. The action of certain 
automorphisms of the group does not descend to a well-defined action on the 
relevant coset space; rather, a single coset finds itself mapped to a family 
of cosets. This ``explains'' the appearance of constants of integration in
the implementation of BTs. I expect this underlying algebraic idea to be 
behind many, if not all, other BTs that involve the solution of differential 
equations. 

These comments are relevant to the discussion of BTs of a single parameter 
value. Another issue to consider is the following: applying BTs 
of two different parameter values $\theta_1,\theta_2$ to a solution gives
two families of solutions. Is there a map between the families? The natural 
guess for such a map would be the map $B(\theta_1)B(\theta_2)^{-1}$ on $G$, 
if it descended to the double coset space, which it does not
This map has already been given
explicitly in Eq.~(\ref{BB}) (since $B(\theta)^{-1}=B(\theta)$); it is
the right multiplication of $G$ by a certain element of $G$, and I will
consider such transformations, including this one, in Sec.\ref{ZSDT}.

Finally in this section, since I have now identified BTs at the level of
the loop group, I can write down the loop group elements corresponding
to single soliton solutions. Applying the BT ${\cal O}_\theta$ to the 
identity in the loop group gives the loop group elements
\begin{equation}
U_0 = \pmatrix{ s\sqrt{\frac{\lambda-\theta}{\lambda}} &
                \frac{r}{\sqrt{\lambda(\lambda-\theta)}} \cr
                q   \sqrt{\lambda(\lambda-\theta)} &
                p\sqrt{\frac{\lambda}{\lambda-\theta}} \cr}.
\label{solU1}\end{equation}
These are loop group elements only for $\vert\theta\vert<1$; they
give the standard soliton solutions
\begin{eqnarray}
 j &=&-t_{-1}-\sqrt{\theta}~{\rm tanh}(z\sqrt{\theta}+A) \label{1sol}\\
          u &=&-\theta~{\rm sech}^2(z\sqrt{\theta}+A), \nonumber
\end{eqnarray}
where 
\begin{equation}
z=\sum_{n=0}^{\infty} t_n\theta^n~~~~~~~~~~
{}~~~~~~e^A=\sqrt{\frac{p\sqrt{\theta}+r}{p\sqrt{\theta}-r}}.
\label{1solpar}\end{equation}
I have assumed that $p\sqrt{\theta}\not=\pm r$; if $p\sqrt{\theta}=\pm r$
the  trivial solutions $j=-t_{-1}\mp\sqrt{\theta}$, $u=0$ emerge.
Note that only soliton solutions with $\vert\theta\vert<1$
arise. This is in accord with the results of \cite{SW}. 

\subsection{Galas B\"acklund Transformations}
\label{GBT}

Galas BTs are, in their algebraic structure, very similar to standard 
BTs. The starting point is the map $C:G\rightarrow G$ defined by
\begin{equation} 
U_0 \mapsto U_0 \pmatrix{1 &0\cr \mu/\lambda& 1\cr}
\end{equation}
where $\mu$ is a nonzero parameter. This descends to $H\backslash G$ but
not to the double coset space, since for $\pmatrix{a&b\cr c&d\cr}\in G^+$,
\begin{equation}
U_0  \pmatrix{a&b\cr c&d\cr} \pmatrix{1 &0\cr \mu/\lambda& 1\cr} =
U_0 \pmatrix{1 &0\cr \mu/\lambda& 1\cr} 
\pmatrix{a+\frac{b\mu}{\lambda} & b\cr c+\frac{(d-a)\mu}{\lambda}
-\frac{b\mu^2}{\lambda^2} & d - \frac{b\mu}{\lambda} \cr} .
\end{equation}
It emerges, though, from this, that $C$ does descend to $H\backslash G/K$,
where $K$ is the codimension two subgroup of $G^+$ consisting of  
matrices $\pmatrix{a&b\cr c&d\cr}$ with $b$ having a zero at $\lambda =0$,
and the constant term in the power series expansion of $d-a-b/\lambda$
also vanishing (a simple calculation is required to check these properties 
indeed define a group). Furthermore, $G^{+(2)}\subset K\subset G^+$, where 
$G^{+(2)}$ is the normal subgroup of $G^+$ consisting of matrices which have
constant term the identity, and no linear term. Thus, defining the map
${\cal P}_0$ via 
\begin{equation}
U_0 \mapsto U_0 \pmatrix{p&q\cr r&s\cr}  
\frac1{\sqrt{1-\lambda^2(P^2+QR)}} 
    \left(I+\lambda\pmatrix{P&Q\cr R&-P\cr}\right)
\pmatrix{1 &0\cr \mu/\lambda& 1\cr},
\end{equation}
where here $\pmatrix{p&q\cr r&s\cr}$ is an arbitrary $SL(2,{\bf C})$ matrix 
and $\pmatrix{P&Q\cr R&-P\cr}$ is an arbitrary $sl(2,{\bf C})$ matrix, gives
a map that will descend to the double coset space: I have inserted after
$U_0$ a group element from each class in $G^+/G^{+(2)}$. It should
be noted that there is no natural isomorphism between  $G^+/G^{+(2)}$
and a subgroup of $G^+$.

Note that 
\begin{equation}
\pmatrix{1 &0\cr \mu/\lambda& 1 \cr}  = \pmatrix{\mu^{-\frac12} &0\cr
  0&\mu^{\frac12}\cr}  \pmatrix{1 &0\cr 1/\lambda & 1\cr}  
  \pmatrix{\mu^{\frac12} &0\cr 0& \mu^{-\frac12}\cr}.
\end{equation}  
Since the first and third of the matrices on the RHS are both constant,
and right multiplication of $G$ by any constant matrix descends to the 
identity on $G/G^+$, it follows that, without any loss of 
generality I can set $\mu=1$, which I do from here on. 

Before computing the effect of ${\cal P}_0$ on a KdV solution, I 
enlarge to a one parameter family of maps in the same way as in 
Sec.~\ref{WEBT}, defining 
\begin{eqnarray}
C(\theta)&=&A(-\theta)~B~A(\theta) \\
C(\theta) U_0 &=& U_0 \pmatrix{1&0\cr \frac1{\lambda-\theta}&1\cr},
   \nonumber\\
{\cal P}_\theta&=&A(-\theta)~{\cal P}_0~A(\theta) \\
{\cal P}_\theta U_0 &=& U_0 \pmatrix{p&q\cr r&s\cr}
\frac1{\sqrt{1-(\lambda-\theta)^2(P^2+QR)}}  \nonumber\\
  &~&  \left(I+(\lambda-\theta)\pmatrix{P&Q\cr R&-P\cr}\right)
 \pmatrix{1&0\cr \frac1{\lambda-\theta}&1\cr}.\nonumber 
\end{eqnarray}
Note that $C(\theta_1)C(\theta_2)=C(\theta_2)C(\theta_1)$, i.e. here
the relevant maps on $G$ are commutative. It is not immediately
obvious that this implies ${\cal P}_{\theta_1}{\cal P}_{\theta_2}=
{\cal P}_{\theta_2}{\cal P}_{\theta_1}$; the simplest way to be 
satisfied that this is indeed the case is to check that if the matrix 
${\cal M}$ is defined by 
\begin{eqnarray}
& \pmatrix{a_1&b_1\cr c_1&d_1\cr}
\pmatrix{1&0\cr \frac1{\lambda-\theta_1}&1\cr}
\pmatrix{a_2&b_2\cr c_2&d_2\cr}
\pmatrix{1&0\cr \frac1{\lambda-\theta_2}&1\cr}=& \\
& {\cal M}~ \pmatrix{1&0\cr \frac1{\lambda-\theta_2}&1\cr}
\pmatrix{a_1&b_1\cr c_1&d_1\cr}
\pmatrix{1&0\cr \frac1{\lambda-\theta_1}&1\cr}, &\nonumber
\end{eqnarray}
with $\pmatrix{a_1&b_1\cr c_1&d_1\cr},\pmatrix{a_2&b_2\cr c_2&d_2\cr}
\in G^+$, then ${\cal M}\in G^+$.

To obtain the effect of ${\cal P}_{\theta}$ on a KdV solution,  observe 
that under ${\cal P}_{\theta}$
\begin{eqnarray}
U &\mapsto& U \pmatrix{p&q\cr r&s\cr}  \nonumber\\
&~&\frac1{\sqrt{1-(\lambda-\theta)^2(P^2+QR)}} 
    \left(I+(\lambda-\theta)\pmatrix{P&Q\cr R&-P\cr}\right)
     \pmatrix{1&0\cr \frac1{\lambda-\theta}&1\cr}  \nonumber\\
  &=& S^{-1} \left( I - \frac{N}{\lambda-\theta}\right)
     \left( I + \frac{N}{\lambda-\theta}\right) Y   \pmatrix{p&q\cr r&s\cr}
     \label{split}\\
  &~&\frac1{\sqrt{1-(\lambda-\theta)^2(P^2+QR)}} 
    \left(I+(\lambda-\theta)\pmatrix{P&Q\cr R&-P\cr}\right)
\pmatrix{1&0\cr \frac1{\lambda-\theta}&1\cr},     \nonumber
\end{eqnarray}
where in the last line I have inserted a certain factor and its inverse,
$N$ being a nonzero matrix independent of $\lambda$ with $N^2=0$. A 
tedious calculation shows that it is possible to choose such a 
matrix $N$ so that the above is written in Birkhoff factorized form.
To write it requires a little more notation.
Referring back to Eqs.~(\ref{Yth1})-(\ref{Yth2}), I write
\begin{equation}
(Y_0^{\theta})^{-1}Y_1^\theta = 
\pmatrix{b_1^\theta & b_2^\theta \cr b_3^\theta & -b_1^\theta \cr}.
\end{equation}
Then correct choice of $N$ is given by
\begin{eqnarray}
N &=& \frac1{1+Q+2sqb_1^\theta+s^2b_2^\theta-q^2b_3^\theta} \\
&~&\pmatrix{-(qa_1^\theta+sa_2^\theta)(qa_3^\theta+sa_4^\theta) &
         (qa_1^\theta+sa_2^\theta)^2 \cr
         -(qa_3^\theta+sa_4^\theta)^2 &
         (qa_1^\theta+sa_2^\theta)(qa_3^\theta+sa_4^\theta)\cr}.
         \nonumber
\end{eqnarray}
Using 
\begin{equation}
S \mapsto \left(1+\frac{N}{\lambda-\theta}\right) S
\end{equation}
and formulae for the $t_0$ and $t_1$ derivatives of the $b_i^\theta$ which 
are easily computed, produces, after some labor,  the Galas BT
\begin{equation}
u \mapsto u - \left(\frac{\tau_{t_0}}{\tau}\right)_{t_0}  
{}~~~~~~~~~~~j \mapsto j - \frac{\tau_{t_0}}{\tau}
\end{equation}
where $\tau=1+Q+2sqb_1^\theta+s^2b_2^\theta-q^2b_3^\theta$
is related to $u$ by $(2u+\theta)=(\sqrt{\tau_{t_0}})_{t_0t_0}/
\linebreak[0]\sqrt{\tau_{t_0}}$, and 
\begin{equation}
\frac{\tau_{t_1}}{\tau_{t_0}} =  
\frac{3\theta}2 + \frac14\left(\frac{\tau_{t_0t_0t_0}}
    {\tau_{t_0}}-\frac{3\tau_{t_0t_0}^2}{2\tau_{t_0}^2} \right)
\end{equation}
(c.f. Eqs.~(\ref{GBT1})-(\ref{GBT2}) above). This is, of course, a BT
of the entire hierarchy; further $t_i$ derivatives of $\tau$ can be
computed as desired.

One difference between the Galas BT and standard BTs is that since
\begin{equation}
\pmatrix{1&0\cr 1/(\lambda-\theta) &1\cr}=
\exp\pmatrix{0&0\cr 1/(\lambda-\theta) &0\cr},
\end{equation}
an infinitesimal generator can be written for Galas BTs, as discovered
by Galas \cite{Galas}. Galas BTs do not, on the other hand, have the
involutiveness property of standard BTs. 

It just remains to make some further comments about soliton solutions.
As mentioned in Sec.~\ref{Galas}, Galas BTs can be used to obtain
standard soliton solutions, as well as more general solutions. Thus,
as in Sec.~\ref{WEBT}, the results of this section give a way to find 
matrices $U_0$ corresponding to soliton solutions. Following through
the necessary calculations, it emerges that the matrix
\begin{equation}
U_0 = \pmatrix{p&q\cr p\sqrt{\theta}-\frac1{q} & q\sqrt{\theta}}
      \pmatrix{1 & (\lambda-\theta)Q \cr 0&1\cr}
      \pmatrix{1&0\cr \frac1{\lambda-\theta} &1\cr}~~~~~~~~q\not=0
\label{solU2}\end{equation}
gives the soliton solution 
\begin{eqnarray}
 j &=&-t_{-1}-\sqrt{\theta}~\left(1+ {\rm tanh}(z\sqrt{\theta}+A) \right)
      \\
 u &=&-\theta~{\rm sech}^2(z\sqrt{\theta}+A), \nonumber
\end{eqnarray}
(c.f. Eq.~(\ref{1sol})), with $z$ defined as in Eq.~(\ref{1solpar}) 
and $A$ determined by
\begin{equation}
e^{-2A} = 1 + \frac{2(1+Q)}{q^2}.
\end{equation}
I have not, as of yet, succeeded in determining whether this matrix $U_0$
is (up to a translation in $t_{-1}$) a representative of the same double
coset as the matrix $U_0$ given in Eq.(\ref{solU1}) (after appropriate
adjustment of parameters). This begs the 
general question of whether the map from $H\backslash G/G^+$ to solutions of 
KdV is 1-1, which is also not considered here. The matrix $U_0$ given in
Eq.~(\ref{solU2}) is in $G$ for all $\theta$ with $\vert\theta\vert\not=1$,
and for $\vert\theta\vert>1$ it is in $G^+$;
in performing the Birkhoff decomposition to obtain the soliton solution
I have assumed $\vert\theta\vert<1$.

\subsection{Zakharov-Shabat Dressing Transformations}
\label{ZSDT}
I now consider general right multiplications on $G$, i.e. 
transformations of the form
\begin{equation}
U_0 \mapsto U_0 \cdot g~~~~~~~~~~~~~~g\in G.
\end{equation}
Galas BTs are of this form, and indeed all ``symmetry actions'' can be
rewritten in this form (e.g. a left multiplication $U_0\mapsto g\cdot U_0$ 
can be rewritten $U_0\mapsto U_0\cdot(U_0^{-1}gU_0)$), but here I only want 
to consider field independent transformations, i.e. the case where the 
matrix $g$ is not dependent on $U_0$. Now, such transformations evidently
descend to $H\backslash G$, but in general will not descend to the 
double coset space. But let us for the moment ignore this.
Under such a transformation $U\mapsto U\cdot g = \linebreak[0] 
S^{-1}\cdot Y \cdot g=\linebreak[0] S^{-1} \cdot (YgY^{-1}) \cdot Y$, so 
writing the Birkhoff decomposition for $YgY^{-1}$ in the form $YgY^{-1}=
(YgY^{-1})_-^{-1}(YgY^{-1})_+$ gives the result
$Y\mapsto (YgY^{-1})_+Y=(YgY^{-1})_-Yg$. This is almost exactly 
the formula for dressing transformations given by Wilson \cite{Wilson};
Wilson's formula actually reads $Y\mapsto (YgY^{-1})_-^{-1}Y$, and 
differs from the one given here by right multiplication by a factor of $g$,
which does not contribute to $dY\cdot Y^{-1}$. Thus the transformations 
being considered in this section are Zakharov-Shabat dressing 
transformations. As explained in Sec.~\ref{dress}, these transformations
are only well-defined  on $Y$; given a solution of the KdV hierarchy,
there are many ways to construct an appropriate $Y$, and dressing 
transformations do not preserve equivalence classes of $Y$'s. This is 
precisely the issue that right multiplications do not in general descend
to $H\backslash G/G^+$. 

In the case of Galas BTs I have shown how to use a right multiplication that
does not descend to  $H\backslash G/G^+$ to obtain a BT of the hierarchy,
a map that takes single solutions to a finite dimensional family of
solutions. It is clear that the procedure followed can be replicated 
whenever the right multiplication descends to a double coset space
$H\backslash G/K$ where now $K$ is any subgroup of finite codimension in 
$G^+$; in particular this can be done whenever $g$ is a rational element of
$G$, i.e. has entries which are rational functions of $\lambda$, with no
singularities on $\vert\lambda\vert=1$. It would  be interesting to 
find a set of generators for the group of rational loops and to determine
the associated BTs; though BTs can, of course, also arise from 
transformations other than right multiplications. 

The problem that dressing transformations suffer from,
viz. the fact that they cannot be really considered as transformations on
the space of solutions of the hierarchy, can be resolved by specifying
a choice of $Y$ or $U_0$ corresponding to a solution of the hierarchy.
For the case of the trivial solution of the hierarchy it is natural to
choose $U_0$ to be the identity matrix, giving (at least for $t_{-1}=0$) 
$Y=M$. This choice was made by Wilson \cite{Wilson}, and it allowed him to
define the orbit of the trivial solution of KdV under dressing 
transformations. 

An important context in which dressing transformations arise is 
understanding the relationship between modified KdV (MKdV)
flows and Liouville and sine-Gordon flows. So far I have not 
constructed MKdV flows in this paper. In fact, associated with any 
solution of
Eq.~(\ref{Uequn}), the fundamental linear equation underlying the KdV
system, there are two MKdV fields, given, in the notation of 
Eq.~(\ref{Ycpts}), by
\begin{equation}
v_1 = {{a_{1t_0}}\over{a_1}}~~~~~~~~{\rm and}~~~~~~~~
v_2 = {{a_{2t_0}}\over{a_2}}.
\end{equation}
{}From Eq.~(\ref{strucequns}), these are related to the KdV field $u$ 
by the so-called Miura map
\begin{equation}
2u = v_{1t_0} + v_1^2  = v_{2t_0} + v_2^2 ,
\end{equation}
and, from Eq.~(\ref{t1els}), they satisfy the MKdV equation
\begin{eqnarray}
v_{1t_1} &=& {\textstyle{1\over 4}}v_{1t_0t_0t_0}-       
             {\textstyle{3\over 2}}v_1^2v_{1t_0} \nonumber\\
v_{2t_1} &=& {\textstyle{1\over 4}}v_{2t_0t_0t_0}-       
             {\textstyle{3\over 2}}v_2^2v_{2t_0}.
\end{eqnarray}
Indeed $v_1,v_2$ satisfy the full MKdV hierarchy. Let us now introduce 
two  new flows for $U$, given by 
\begin{eqnarray}
\partial_{s_1} U &=& U \pmatrix{0&1\cr 0&0\cr}   \\
\partial_{s_2} U &=& U \pmatrix{0&0\cr \lambda^{-1}&0\cr}. \label{s2equn}
\end{eqnarray}
The first of these flows corresponds to an infinitesimal
dressing transformation by  a constant matrix, and hence leaves the KdV
field $u$ invariant. It is also straightforward to check that this
flow leaves the field $a_1$ invariant, and hence also $v_1$, but induces
a nontrivial flow on $v_2$ given by
\begin{equation}
v_{2s_1}=-{1\over{a_2^2}}.
\end{equation}
Since $v_2=a_{2t_0}/a_2$, by introducing $h=\log a_2$, this
takes the form of the Liouville equation $h_{t_0s_1}=-e^{-2h}$. The flow
in Eq.~(\ref{s2equn}), is an infinitesimal Galas transformation, and does
not leave $u$ invariant. The effect on $a_1,a_2$ can be computed ---
with some effort --- from the Birkhoff decomposition Eq.~(\ref{split}),
and the relevant fact for the current discussion is that 
\begin{equation}
v_{2s_2}=a_2^2,
\end{equation}
also a Liouville equation. A suitable linear combination of the 
$s_1$ and $s_2$ flows gives the sine-Gordon (or, more properly,
the sinh-Gordon) equation. Both the $s_1$ and $s_2$ evolutions
for $v_2$ commute with all the MKdV flows; this is obvious 
because the MKdV flows are implemented by left 
multiplications on $U_0$, and the $s_1$ and $s_2$ flows by right 
multiplication. However, the $s_1$ and $s_2$ flows do not commute with 
each other. 

The MKdV equations, and Liouville and sine-Gordon flows will not be 
discussed in full detail here, but a few more notes are in order. First,
that both $s_1$ and $s_2$ flows for $v_2$ commute with the MKdV flow is
a symptom of the symmetry of the MKdV hierarchy under the involution 
$v_2\mapsto -v_2$, which can be shown to be an effect of the involution
$B$ used in the discussion of Wahlquist-Estabrook BTs. Second, the fact
that the MKdV fields $v_1,v_2$ are really ``derived'' quantities, and the
centrality of the fields $a_1,a_2$ has been stressed in \cite{oldWil}
(see also \cite{Mcandme}). Finally, a detailed studied of BTs of the 
sine-Gordon system has been given in \cite{Uhl}, and there is much in
common between the results of \cite{Uhl} and the current work.

Finally in this section, I reconsider briefly the dressing transformation 
introduced in Sec.~\ref{WEBT} as the product of two 
Wahlquist-Estabrook BTs, Eq.~(\ref{BB}).
The remarkable fact about this dressing transformation is that despite
the fact that it is not multiplication by a rational element of $G$, it
still has an interpretation as a BT. For completeness I write down here
the infinitesimal version of this transformation (taking $\theta_2=0$ and
$\theta_1=2\epsilon$, where $\epsilon$ is an infinitesimal parameter). 
This is then the dressing transformation
\begin{equation}
U_0 \mapsto U_0 \pmatrix{ 1+\epsilon/\lambda & 0\cr
                       0 & 1-\epsilon/\lambda \cr}.
\end{equation}
Following through the necessary calculations, it an be shown that 
this induces the infinitesimal BT 
\begin{eqnarray}
j &\mapsto& j -2\epsilon (pa_1+ra_2)(qa_1+sa_2)  \\
u &\mapsto& u -2\epsilon \left((pa_1+ra_2)(qa_1+sa_2)\right)_{t_0}, 
\end{eqnarray}
where, as usual, $\pmatrix{p&q\cr r&s\cr}$ is an arbitrary $SL(2,{\bf C})$
matrix, and $a_1,a_2$ are as above.
Since the variation of $j$ and the determinant condition $ps-rq=1$
are invariant under the rescalings 
$p\mapsto fp,~r\mapsto fr,~q\mapsto f^{-1}q, s\mapsto f^{-1}s$, 
$f\in{\bf C}^*$, this infinitesimal map generates a two parameter family of 
solutions from a given one, which is consistent with its origins in 
Sec.~\ref{WEBT}. Looking at this formula alone, there is little reason
to suspect that this infinitesimal symmetry can be exponentiated to
a finite BT.

\section{Open Problems}
\label{concl}
Some open problems have already been mentioned in the text, and will
not be repeated. The
central open issue that I perceive is that although in this paper 
a reasonable unified framework has been built for understanding symmetries
of KdV, this has been in the context of a restricted set of solutions.
It is quite possible that there is a simple modification of the 
formalism presented here that will allow consideration of much more 
general spaces of solutions (c.f. \cite{Mulase}). 

Another issue that has not been tackled in this paper is that I have
not discussed the Hamiltonian structures or the conserved quantities of
the KdV hierarchy. There is, of course, in general, a connection between
symmetries and conserved quantities, and it is of considerable interest
to understand what the conserved quantities (or appropriate generalizations
thereof) associated to BTs are. For that matter, while the conserved 
quantities associated with the translation symmetries of KdV are of
course well known, I am unaware of whether the conserved quantities
associated to Galilean or scaling symmetries have been discussed.

It is to be hoped that the work presented here will be continued in several
directions. The study of KdV presented here is based on two hypotheses,
Mulase's belief that beneath each integrable system lies a genuinely 
simple linear system, like Eq.~(\ref{Uequn}), and the idea that the most
logically satisfactory explanation of BTs is that they arise from simple
actions on the space of initial data of the relevant simple linear system.
These ideas need to be explored for other integrable systems. Initial studies 
for both the KP and Principal Chiral Model hierarchies \cite{ip} confirm this 
general picture. Far from being an esoteric study of the structure of 
integrable systems, this program holds promise of the discovery of new BTs
of integrable systems, and, through this, an expansion of the repertoire
of known solutions. Indeed, I originally found the Galas BT for  $\theta=0$ 
via the considerations of Sec.~\ref{GBT}, before finding Galas' paper 
\cite{Galas}, and there seems no reason why the Galas BT should not occupy 
just as significant a place in the textbooks on soliton equations as the 
Wahlquist-Estabrook BT. 

{}From this paper also emerges the need for further studies and 
classifications of the solutions of KdV, particularly a deeper exploration 
of the space of solutions that has been considered here, and reconciliation 
with the zoo of solutions available in the literature \cite{ratsol}.

\section*{Acknowledgments} I thank Steve Shnider for much encouragement
and many useful suggestions, Peter Olver for a few critical comments,
and the Rashi Foundation for support via
a Guastella Fellowship.

\section*{Appendix}
I complete here the proof of the theorem from Sec.~\ref{zc}. 
Eqs.~(\ref{curvs}) clearly
imply, for $m\ge 0$, $n\ge -1$, that
\begin{equation}
\frac{\partial (Z_m-\lambda Z_{m-1})}{\partial t_n}-
\left( \frac{\partial}{\partial t_m}-\lambda
    \frac{\partial}{\partial t_{m-1}} 
                             \right)Z_n
      + [Z_m-\lambda Z_{m-1},Z_n]  = 0.
\label{a1}\end{equation} 
Using $Z_m-\lambda Z_{m-1}=\Delta_m$ and $Z_n=\sum_{i=0}^{n+1}\Delta_{n-i}
\lambda^i$ (where, following the convention of Sec.~\ref{the-constr}, 
$\Delta_{-1}=\pmatrix{0&0\cr 1&0\cr}$), and equating powers of $\lambda$,
Eq.~(\ref{a1}) gives the following equations:
\begin{eqnarray}
\frac{\partial \Delta_0}{\partial t_{m-1}}+[\Delta_m,\Delta_{-1}] &=& 0 
{}   ~~~~~~~~~~m\ge 0 \label{a2} \\
\frac{\partial\Delta_{n+1}}{\partial t_{m-1}}-
   \frac{\partial\Delta_{n}}{\partial t_{m}} + [\Delta_m,\Delta_{n}]
   &=& 0 ~~~~~~~~~~ m,n \ge 0 \label{a3} \\
\frac{\partial\Delta_{m}}{\partial t_{n}} -
   \frac{\partial\Delta_{n}}{\partial t_{m}} +
   [\Delta_m,\Delta_n]& =& 0 ~~~~~~~~~~m \ge 0,~n\ge -1 \label{a4}
\end{eqnarray}
The first set of these reproduces Eq.~(\ref{what}). 
Note the other two sets imply the interesting result
\begin{equation}
\frac{\partial\Delta_{n+1}}{\partial t_{m-1}}=
\frac{\partial\Delta_{m}}{\partial t_{n}}~~~~~~~~m,n \ge 0.
\label{aa5}\end{equation}

Using Eqs.~(\ref{bitdel}), the 1,2 entry of Eq.~(\ref{a3}) 
or Eq.~(\ref{a4}) gives
\begin{equation}
\left(\frac{\partial}{\partial t_{m}} \frac{\partial}{\partial t_{n-1}} -
      \frac{\partial}{\partial t_{n}} \frac{\partial}{\partial t_{m-1}}
\right)j = 
  \frac{\partial u}{\partial t_{m-1}} \frac{\partial j}{\partial t_{n-1}} 
 -\frac{\partial u}{\partial t_{n-1}} \frac{\partial j}{\partial t_{m-1}}
\label{a5}\end{equation}
The situation faced at the end of Sec.~\ref{zc}
was that given the correct $t_{n-1}$ flow for $j,u$ ($n\ge 1$),
the correct $t_n$-flow for $u$ could be deduced, but this was not sufficient
to uniquely determine the correct $t_n$-flow for $j$, and instead gave us
\begin{equation}
\frac{\partial j}{\partial t_n} = 
({\textstyle{\frac14}}\partial_{t_0}^2 - \partial_{t_0}^{-1} u\partial_{t_0} 
  - u)
\frac{\partial j}{\partial t_{n-1}} + c_n(t_{1},t_{2},\ldots).
\label{a6}\end{equation}
(Here I have integrated both sides of Eq.~(\ref{nearly}). The integration
operator $\partial_{t_0}^{-1}$ can be precisely defined, see \cite{Olver}.)
Differentiating with respect to $t_{m-1}$, Eq.~(\ref{a6}) gives 
\begin{equation}
\frac{\partial^2 j}{\partial t_n\partial t_{m-1}} = 
- \frac{\partial u}{\partial t_{m-1}}
\frac{\partial j}{\partial t_{n-1}} + 
\frac{\partial c_n}{\partial t_{m-1}} +
\pmatrix{{\rm terms~symmetric}\cr{\rm in}~~m,n\cr} .
\label{a7}\end{equation}
This, with Eq.~(\ref{a5}), implies
\begin{equation}  
\frac{\partial c_n}{\partial t_{m-1}}=
\frac{\partial c_m}{\partial t_{n-1}}
{}~~~~~~~~m\ge 0,~n\ge 1.
\label{a8}\end{equation}   
As an application of this, set $n=1$ to obtain that $c_1$ is
constant (and therefore, by an admissible change of coordinates, can be
set to zero). However, it is clear that Eq.~(\ref{a8}) by itself is 
not enough to eliminate the $t_n$ dependence from all the $c_n$'s
(it admits solutions of the form $c_n=\partial P/\partial t_{n-1}$ for
appropriate functions $P$).
To do this, it is necessary
to look at further entries of Eqs.~(\ref{a3})-(\ref{a4});
in particular looking the 2,1 entry of Eq.~(\ref{a3}), and using 
\begin{equation}
(\Delta_n)_{21}= -c_n +
\left({\textstyle{1\over 4}}\partial_{t_0}^2-\partial_{t_0} j
   + j^2 +\partial_{t_0}^{-1} u \partial_{t_0} \right)
   {{\partial j}\over{\partial_{t_{n-1}}}},
\end{equation}
which follows from Eq.(\ref{mad1}) and Eq.(\ref{a6}), gives, after 
some manipulations, the further requirement
\begin{equation}
{{\partial c_n}\over{\partial_{t_m}}}=
{{\partial c_m}\over{\partial_{t_n}}}.
\end{equation}
This with Eq.~(\ref{a8}) is sufficient to imply that all the $c_n$'s
are constant.


\begin{thebibliography}{99}
\bibitem{ratsol} Ablowitz, M.J., Cornille,H.:
  On solutions of the Korteweg-De Vreis Equation. 
  Phys.Lett.A {\bf 72}, 277-280 (1979).
  Matveev, V.B.: Generalized Wronskian formula for solutions of the KdV
  equations: first applications.
  Phys.Lett.A {\bf 166}  205-208 (1992) (erratum, {\em ibid.}
  {\bf 168} 463 (1993)). 
  Matveev, V.B.: Positon-positon and soliton-positon collisions: KdV case.
  Phys.Lett.A {\bf 166}  209-212 (1992).
  Matveev, V.B.: Asymptotics of the multipositon-soliton $\tau$ function
  of the Korteweg-De Vries equation and the supertransparency.
  J.Math.Phys. {\bf 35} 2955-2970 (1994). 
  Rasinariu, C., Sukhatme, U., Khare, A.: 
  Negaton and Positon Solutions of the KdV Equation.
  Preprint, hep-th/9505133.
\bibitem{ASvM} Adler, M., Shiota, T., van Moerbeke, P.:
  A Lax Representation for the Vertex Operator and the Central Extension.
  Comm.Math.Phys. {\bf 171} 547-588 (1995).
\bibitem{CLL} Chen, H.H., Lee, Y.C., Lin, J.E.: On a New Hierarchy of 
   Symmetries for the Kadomtsev-Petviashvili Equation.
 Physica D {\bf 9} 439-445 (1983).
\bibitem{DJ} Drazin, P.G. and Johnson, R.S.:  Solitons: an Introduction.
   Cambridge University Press 1989.
\bibitem{DS} Drinfeld, V.G., Sokolov, V.V.: Lie Algebras and Equations 
  of Korteweg-De Vries Type.
  {\em J.Sov.Math.} {\bf 30} 1975-2036  (1985).
\bibitem{Galas} Galas, F.: New non-local symmetries with psuedopotentials.
   J.Phys.A: Math.Gen. {\bf 25} L981-L986 (1992).
\bibitem{GeS} Gesztesy, F., Simon, B.: Constructing Solutions of 
  the mKdV-Equation. J.Func.Anal. {\bf 89}  53-60 (1990).
\bibitem{GH1} Guthrie, G.A.: More non-local symmetries of the KdV equation.
 J.Phys.A: Math.Gen. {\bf 26}  L905-L908 (1993).
\bibitem{GH2} Guthrie, G.A., Hickman, M.S.: Nonlocal symmetries of the KdV
  equation. J.Math.Phys. {\bf 34} 193-205 (1993).
\bibitem{HSS} Haak, G., Schmidt, M., Schrader, R.: Group Theoretic
 Formulation of the Segal-Wilson Approach to Integrable Systems with 
 Applications.  Rev.Math.Phys. {\bf 4} 451-499 (1992).
\bibitem{IS} Ibragimov, N.Kh.,  Shabat, A.B.: the Korteweg-de Vries equation
  from the point of view of transformation groups.
  Sov.Phys.Dokl. {\bf 24} 15-17 (1979).
\bibitem{Kor} Khor'kova, N.G.: Conservation Laws and Nonlocal Symmetries.
 Math.Notes {\bf 44} 562-568 (1989).
\bibitem{Mcandme} McIntosh, I.: ${\rm SL(n+1)}$-invariant equations which
  reduce to equations of Korteweg-de Vries type.
 Proc.Roy.Soc.Edin. {\bf 115A} 367-381 (1990).
 Depireux, D., Schiff, J.: On UrKdV and UrKP.
 Lett.Math.Phys. {\bf 33} 99-111 (1995).
\bibitem{Mulase} Mulase, M.: Complete Integrability of the 
   Kadomtsev-Petviashvili Equation. Adv.Math. {\bf 54} 57-66 (1984).
  Mulase, M.: Solvability of the super KP equation and a generalization 
  of the Birkhoff decomposition. Inv.Math. {\bf 92} 1-46 (1988).
\bibitem{hidsym} Oevel, G., Fuchssteiner, B., Blaszak, M.:
Action-Angle Representation of Multisolitons by Potentials of Mastersymmtries.
Prog.Theo.Phys. {\bf 83} 395-413 (1990).
\bibitem{Olver} Olver, P.J.: Applications of Lie Groups to
  Differential Equations. Springer-Verlag 1986.
\bibitem{loopgroups} Pressley, A., Segal, G.: Loop Groups.
    Oxford: Clarendon Press 1986.
\bibitem{ip} Schiff, J., in preparation. 
\bibitem{SW} Segal, G., Wilson, G.: Loop Groups and Equations of KdV Type.
  Pub.Math.I.H.E.S {\bf 61} 5-65 (1983).
\bibitem{Uhl} Uhlenbeck, K.: On the connection between harmonic maps and
  the self-dual Yang-Mills and the Sine-Gordon equations.
 J.Geom.Phys. {\bf 8} 283-316 (1992).
\bibitem{WE} Wahlquist, H.D., Estabrook, F.B.: B\"acklund Transformations
  for Solutions of the Korteweg-de Vries Equation. Phys.Rev.Lett. {\bf 31} 
  1386-1390 (1973).
\bibitem{Wilson} Wilson, G.: Habillage et fonctions $\tau$.
  C.R. Acad.Sc.Paris {\bf 299} 587-590 (1984). Wilson, G.:
  Infinite-dimensional Lie groups and algebraic geometry in soliton theory.
  Phil.\linebreak[0]Trans.\linebreak[0]Roy.\linebreak[0]Soc.\linebreak[0]Lond.
  {\bf 315}  393-404 (1985).
\bibitem{oldWil} Wilson, G.: On the Quasi-Hamiltonian Formalism 
  of the KdV Equation.  Phys.Lett.A {\bf 132}  445-450 (1988).

\end{thebibliography}
\end{document}